\documentclass[eqsecnum,showpacs,aps]{revtex4}

\usepackage{color}

\usepackage{graphicx}
\usepackage{amsmath,amssymb}
\usepackage{bm}

\newcommand{\beq}{\begin{equation}}
\newcommand{\eeq}{\end{equation}}

\def \({\left( }
\def \){\right) }
\newcommand{\beqar}{\begin{eqnarray}}
\newcommand{\eeqar}{\end{eqnarray}}

\newcommand{\I}{\mbox{\rm i}}

\newcommand{\dalm}{\kern1pt\vbox{\hrule height 0.9pt\hbox{\vrule width 0.9pt\hskip 2.5pt\vbox{\vskip 5.5pt}\hskip 3pt\vrule width 0.3pt}\hrule height 0.3pt}\kern1pt}

\begin{document}



\title{Gravitationally Driven Electromagnetic Perturbations of\\
 Neutron Stars and Black Holes}

\author{Hajime Sotani$^{1}$, Kostas D. Kokkotas$^{2,3}$,  Pablo Laguna$^{4}$, Carlos F. Sopuerta$^{5}$}
\affiliation{
$^{1}$Yukawa Institute for Theoretical Physics, Kyoto University, Kyoto 606-8502, Japan\\
$^{2}$Theoretical Astrophysics, University of T\"ubingen, IAAT, Auf der Morgenstelle 10, 72076, T\"ubingen, Germany\\
$^{3}$Department of Physics, Aristotle University of Thessaloniki, 54124, Thessaloniki, Greece\\
$^{4}$Center for Relativistic Astrophysics and School of Physics, Georgia Institute of Technology, Atlanta, GA 30332, USA\\
$^{5}$ Institut de Ci\`encies de l'Espai (CSIC-IEEC), Facultat de Ci\`encies,
Campus UAB, Torre C5 parells, Bellaterra, 08193 Barcelona, Spain.
}

\date{\today}

\begin{abstract}
Gravitational perturbations of neutron stars and black holes are well known sources of gravitational radiation.
If the compact object is immersed in or endowed with a magnetic field, the gravitational perturbations would couple
to electromagnetic perturbations and potentially trigger synergistic electromagnetic signatures.
We present a detailed analytic calculation of the dynamics of coupled gravitational and electromagnetic perturbations for both neutron stars and black holes.
We discuss the prospects for detecting the electromagnetic waves in these scenarios and the potential that these waves have for providing information
about their source. 
\end{abstract}

%



\maketitle


\section{Introduction}
\label{sec:Intro}

Multi-messenger astronomy has arrived. Already astro-particle observations (neutrinos and cosmic rays)  are complementing traditional electromagnetic observations.
The third pillar is almost ready with near future gravitational-wave observations by interferometric detectors like LIGO, Virgo, GEO600 and LCGT~\cite{LIGO1, LIGO2}.
This new astronomy will enable multi-channel observations of astrophysical phenomena such as $\gamma$-ray bursts, supernovae, or flaring magnetars, unveiling an unprecedented view of the nature of the source and its environment.

An important component in many astrophysical phenomena is strong magnetic fields, as demonstrated by the active role they play in the accretion processes 
of low-mass X-ray binaries and GRBs \cite{Ghosh}. The presence of strong magnetic fields opens up the possibility for interesting effects. Among them, which is the central 
topic of this work, is the coupling between electromagnetic and gravitational emissions that could yield synergistic multi-messenger observations. 
In particular, it is important to assess the conditions in which electromagnetic and gravitational emissions influence each other.
There are already hints for such scenario. It is believed that the flare activity 
of magnetars seems to be associated with starquakes \cite{WS2006}. These quakes are responsible not only for dramatic 
perturbations and rearrangements of the magnetic field, but also for the breaking of the neutron star crust and 
internal motions, possibly resulting in the emission of 
gravitational waves. Detailed studies of magnetar flare activity have revealed a number of features in the 
afterglow, which can be associated with the crust oscillations as well as with Alfv\'en waves propagating from 
the core towards the 
surface~
\cite{GSA2006,L2007,SKS2007,SKS2008,CBK2009,CSF2009,S2011,HoLevin11,GCFES2012,CK2012,SNIO2012a,SNIO2012b,HoLevin12}. 

The link or coupling between electromagnetic radiation 
and gravitational waves have been investigated for some cases. One of them looked at  the propagation of  gravitational waves 
linearly coupled to an external magnetic field~\cite{GP1980}. It was shown that 
this configuration triggers magneto-hydrodynamics waves in the plasma \cite{PSVK2001, SB2003, MK2003, MK2004,FBMSM2006}.
Furthermore, the linear nature of the coupling limits the electromagnetic waves to low frequencies, in the best case a few tenths of kHz, which will 
be easily absorbed by the interstellar medium or plasma. In order to produce high frequency and detectable electromagnetic waves, 
non-linear couplings are needed, requiring much stronger gravitational waves. In most of these studies, the 
gravitational waves were assumed to propagate on a flat space-time background. This is a reasonable assumption
when the interaction between the gravitational and electromagnetic waves takes place far from the source. There have been only very few attempts to treat the electromagnetic-gravity coupling in the strong field regime~\cite{CMBD2004}.  

The aim of this work is to study the interaction of electromagnetic and gravitational 
waves in the vicinity of magnetized neutron stars or black holes immersed in strong magnetic fields using perturbation theory, paying particular attention to
how gravitational modes drive the excitation of electromagnetic perturbations. Our work also includes estimates of
the energy transferred between the gravitational and electromagnetic sectors. 
As expected, we find that the excited electromagnetic waves have roughly the same frequency 
as the driving gravitational waves, i.e., of the order of a few kHz. Electromagnetic waves at these low frequencies can be easily absorbed by the interstellar medium. As a consequence, one needs to associate them with secondary emission 
mechanisms (e.g., synchrotron radiation) in order to be able to trace the effects of gravitational waves on the strong magnetic fields.  The later process can be studied following 
the mechanisms described in~\cite{PSVK2001, SB2003, MK2003, MK2004,FBMSM2006}, and there is work in progress for the 
special case of strong gravitational fields.

This article is organized as follows: Section~\ref{sec:Background} gives details of the space-time background configuration. 
In Sec.~\ref{sec:Perturbation}, we review the general form of the perturbation equations, their couplings, 
and the angular dependences of the various types of electromagnetic and gravitational perturbations. 
In Sec.~\ref{sec:Pert_Star}, we reduce the equations to the particular case of dipole electromagnetic perturbations driven by the 
quadrupole gravitational mode for the case of a neutron star background. In Sec.~\ref{sec:Pert_BH}, we do the same as in Sec.~\ref{sec:Pert_Star} but for the case
of a black hole and consider both the case of axial and polar gravitational 
perturbations. In Sec.~\ref{sec:Results} we show numerical results or 
dipole electromagnetic waves driven by quadrupole gravitational waves with axial parity for both neutron stars and 
 black holes. Conclusions are given in Sec.~\ref{sec:conclusion}. 
We adopt geometric units, $c=G=1$, where $c$ and $G$ denote the speed of light and the gravitational 
constant, respectively, and the metric signature is $(-,+,+,+)$.

\section{Equations for the Background}
\label{sec:Background}

The background space-times we are considering (neutron stars and black holes) are governed by the Einstein-Maxwell equations, which read:
\begin{eqnarray}
  G_{\mu \nu} &=& 8\pi \left(T_{\mu \nu}+E_{\mu\nu}\right)\, , \label{eq:Ein} \\
  \left(T^{\mu\nu} + E^{\mu\nu}\right)_{;\nu} &=& 0 \, , \label{eq:Ein2}\\
  {F^{\mu\nu}}_{;\nu} &=& 4\pi J^{\mu} \, , \label{eq:Max1}\\
  F_{\mu \nu, \lambda}+F_{\lambda \mu, \nu} +F_{\nu \lambda, \mu} &=& 0 \,, \label{eq:Max2}
\end{eqnarray}
The tensors that appear in these equations are: The Einstein tensor $G_{\mu\nu}$, the Faraday antisymmetric tensor $F_{\mu\nu}$, the electromagnetic four-current $J^{\mu}$,
the energy-momentum tensor of the matter fluid $T_{\mu\nu}$, and the energy-momentum tensor of the electromagnetic field is $E_{\mu\nu}$.  The energy-momentum tensors are explicitly given by
\begin{eqnarray}
  T_{\mu\nu} &=& (\rho+p)u_\mu u_\nu+p g_{\mu\nu}\,, \\
  E_{\mu\nu} &=& \frac{1}{4\pi} \left(g^{\rho\sigma}F_{\rho\mu}F_{\sigma\nu}
            -\frac{1}{4} g_{\mu\nu}F_{\rho\sigma}F^{\rho\sigma} \right)\,,
\end{eqnarray}
where $\rho$ stands for the energy-density, $p$ for the pressure, and $u_\mu$ for the four-velocity of the matter fluid.

The presence of a magnetic field could in principle induced deformations to the neutron star or black hole we are considering. However, 
even for astrophysically strong magnetic fields, $B \sim 10^{16} G$, as in the case of magnetars, the energy of the magnetic field ${\cal E}_B$ is much smaller than the 
gravitational energy ${\cal E}_G$, by several orders of magnitude. In fact,
${\cal E}_B/{\cal E}_G \sim 10^{-4}(B/10^{16} {\mbox [G]})^2$. Therefore, in setting up the background space-time metric, one can ignore the magnetic field.
That is, the background metric has the form
\begin{equation}
  ds^2=-e^{\nu}dt^2+e^{\lambda}dr^2 +r^2(d\theta^2+\sin^2\theta d\phi^2)\,, 
  \label{eq:Schw}
\end{equation}
where the functions $\nu(r)$ and $\lambda(r)$, in the interior of a neutron star, are determined by the well-known Tolman-Oppenheimer-Volkoff (TOV) 
equations~(see, e.g.~\cite{BFS85}) and the matter fluid four-velocity $u^{\mu}=(e^{-\nu/2},0,0,0)$. In the exterior of a neutron star, 
and in the case of a black hole, they are determined by the standard Schwarzschild solution: $e^{-\lambda}=e^{\nu}=1-2M/r$.

\subsection{A Dipole Background Magnetic Field: Exterior region}
\label{sec:DipoleField-exterior}

Next, we compute the magnetic field for both the neutron star and the black hole. We consider first the exterior (vacuum) solution.
In this case, the component of Maxwell equations given by Eq. (\ref{eq:Max2}) is automatically satisfied.  
The magnetic field is then obtained by solving the remaining Maxwell equations, Eqs.~(\ref{eq:Max1}), which in vacuum reads
\begin{equation}
  {F^{\mu\nu}}_{;\nu} = 0\,, \label{eq:Max-ex}
\end{equation}
with $F_{\mu\nu} = A_{\nu,\mu} - A_{\mu,\nu}$.
Since the background space-time is static, it is natural to assume that the magnetic field is also static. 
In addition, we require the magnetic field to be axisymmetric and poloidal,
\begin{equation}
\label{eq:bb}
B^{\mu {\rm(ex)}} = \left(0,e^{-\lambda/2}B_1^{\rm (ex)}(r) \cos\theta,e^{-\lambda/2}B_2^{\rm (ex)}(r)\sin\theta,0\right)\,,
\end{equation}
which has a dependence on the polar coordinate, $\theta$.
From the relation between the magnetic field, the matter fluid velocity $u^{\mu}$, and the field strength 
\begin{equation}
  B_{\mu} = \epsilon_{\mu\nu\alpha\beta}u^{\nu}F^{\alpha\beta}/2 = \epsilon_{\mu\nu\alpha\beta}u^{\nu}A^{\alpha,\beta}\,, \label{eq:bfield}
\end{equation}
where $\epsilon_{\mu\nu\alpha\beta}$ is the complete antisymmetric tensor,  determined by the convention $\epsilon_{0123}=\sqrt{-g}$.
It is not difficult to show that the only non-vanishing component of the vector potential $A_{\mu}$ is the $\phi$-component, which we will denote as $A_{\phi}^{\rm (ex)}$.
Therefore, the vacuum Maxwell equation~(\ref{eq:Max-ex})  in the Schwarzschild background becomes
\begin{equation}
  r^2 \frac{\partial}{\partial r} \left[\left(1-\frac{2M}{r}\right)\frac{\partial A_{\phi}^{\rm (ex)}}{\partial r}\right]
    + \sin\theta \frac{\partial}{\partial \theta}
    \left[\frac{1}{\sin\theta}\frac{\partial A_{\phi}^{\rm (ex)}}{\partial \theta}\right]
    = 0\,. \label{eq:Max-ex2}
\end{equation}
Expanding $A_{\phi}^{\rm (ex)}$ in vector spherical harmonics as
\begin{equation}
  A_{\phi}^{\rm (ex)} = a_{l_M}^{\rm (ex)}(r)\sin\theta\,\partial_{\theta}P_{l_M}(\cos\theta)\,, \label{eq:Aphi}
\end{equation}
we rewrite Eq. (\ref{eq:Max-ex2}) as
\begin{equation}
  r^2 \frac{d}{dr} \left[\left(1-\frac{2M}{r}\right)\frac{da_{l_M}^{\rm (ex)}}{dr}\right] - l(l+1)a_{l_M}^{\rm (ex)} = 0\,.
\end{equation}
The solution of this equation for the dipole case ($l_M=1$) has the form~\cite{Wasserman1983}
\begin{equation}
\label{eq:asol}
  a_{1}^{\rm (ex)} = -\frac{3\mu_d}{8M^3}r^2 \left[\ln\left(1-\frac{2M}{r}\right) + \frac{2M}{r} + \frac{2M^2}{r^2}\right]\,,
\end{equation}
where $\mu_d$ is the magnetic dipole moment for an observer at infinity. With the solution of Eq.~(\ref{eq:asol}) and Eq.~(\ref{eq:bfield}), 
the coefficients of the magnetic field in Eq.~(\ref{eq:bb}) are given by:
\begin{eqnarray}
 B_1^{\rm (ex)}(r) &=& \frac{2a_1^{\rm (ex)}}{r^2} = -\frac{3\mu_d}{4M^3} \left[\ln\left(1-\frac{2M}{r}\right) + \frac{2M}{r} + \frac{2M^2}{r^2}\right]\,, 
     \label{eq:B1-ex} \\
 B_2^{\rm (ex)}(r) &=&-\frac{a_{1,r}^{\rm (ex)}}{r^2} =  \frac{3\mu_d}{4M^3r} \left[\ln\left(1-\frac{2M}{r}\right) + \frac{M}{r} + \frac{M}{r-2M}\right]\,.
     \label{eq:B2-ex}
\end{eqnarray}
Notice that in the limit  $r\to \infty$, 
\begin{equation}
  B_1^{\rm (ex)}(r) \approx \frac{2\mu_d }{r^3}\quad \mbox{and} \quad
  B_2^{\rm (ex)}(r) \approx \frac{\mu_d }{r^4}\,.
\end{equation}

\subsection{A Dipole Background Magnetic Field: Interior region}
\label{sec:DipoleField-interior}

We assume that the magnetic field inside the star is also axisymmetric and poloidal, with current $J_{\mu}=(0,0,0,J_{\phi})$~\cite{Bocquet1995,Konno1999}. 
The ideal MHD approximation is also adopted, i.e. infinite conductivity $\sigma$, which leads to 
$E_{\mu}=F_{\mu\nu}u^{\nu}=0$, as follows from the relativistic Ohm's law
\begin{equation}
  F_{\mu\nu} u^{\nu} = \frac{4\pi}{\sigma}\left(J_{\mu} + u_{\mu}J^{\nu}u_{\nu}\right)\,. \label{eq:Ohm}
\end{equation}
Therefore, the vector potential $A_{\mu}$ is similar to that for the exterior magnetic field, i.e. 
$A_{\mu} = (0,0,0,A_{\phi}^{\rm (in)})$. The counterpart  equation to Eq. (\ref{eq:Max-ex2}) but for the interior is 
\begin{equation}
 e^{-\lambda} \frac{\partial^2 A_{\phi}^{\rm (in)}}{\partial r^2}
     + \frac{1}{r^2}\frac{\partial^2 A_{\phi}^{\rm (in)}}{\partial \theta^2}
     + \left(\nu' - \lambda' \right)\frac{e^{-\lambda}}{2}\frac{\partial A_{\phi}^{\rm (in)}}{\partial r}
     - \frac{1}{r^2} \frac{\cos\theta}{\sin\theta}\frac{\partial A_{\phi}^{\rm (in)}}{\partial \theta}
     = -4\pi J_{\phi}\,. \label{eq:Max-in}
\end{equation}
Expanding both, the vector potential $A_{\phi}^{\rm (in)}$ and the current $J_{\phi}$, in vector spherical harmonics, one gets
\begin{eqnarray}
 A_{\phi}^{\rm (in)}(r,\theta) &=& a_{l_M}^{\rm (in)}(r) \sin\theta\, \partial_{\theta}P_{l_M}(\cos\theta)\,, \\
 J_{\phi}(r,\theta)        &=& j_{l_M}(r) \sin\theta\, \partial_{\theta}P_{l_M}(\cos\theta)\,,
\end{eqnarray}
which can be use to rewrite Eq.~(\ref{eq:Max-in}) as
\begin{equation}
 e^{-\lambda} \frac{d^2 a_{l_M}^{\rm (in)}}{dr^2} + \left(\nu' - \lambda' \right) \frac{e^{-\lambda}}{2}
     \frac{d a_{l_M}^{\rm (in)}}{dr} - \frac{l_M(l_M+1)}{r^2} a_{l_M}^{\rm (in)} = -4\pi j_{l_M}\,. \label{eq:Max-in2}
\end{equation}
It is only feasible to obtain numerical solutions to Eq.~(\ref{eq:Max-in2}), even for the dipole case ($l_M=1$), since among other things the coefficients are also computed numerically from the TOV equations.  In addition, 
when prescribing $j_{1}(r)$, it must satisfy an integrability condition 
(see~\cite{Bonazzola1993, Colaiuda2008} for details). We adopt a current with a functional form~\cite{Konno1999}:
\begin{equation}
  j_{1}(r) = f_0r^2(\rho + p)\,,
\end{equation}
where $f_0$ is an arbitrary constant. In addition, we should impose the following regularity condition at center of the neutron star,
\begin{equation}
  a_{1}^{\rm (in)} = \alpha_{c}r^2 + {\cal O}(r^4)\,,
\end{equation}
where $\alpha_c$ is also an arbitrary constant. These arbitrary constants, $f_0$ and $\alpha_c$, are determined by from the matching conditions at the surface of the star, namely that $a_1$ and $a_{1,r}$ are continuous across the stellar surface. 
Finally, once we have the numerical solution for $a_1(r)$, the magnetic field is obtained from
\begin{equation}
B^{\mu {\rm (in)}} = \left(0,e^{-\lambda/2}B_1^{\rm (in)}(r)\cos\theta,
e^{-\lambda/2}B_2^{\rm (in)}(r)\sin\theta,0\right)
\end{equation}
with
\begin{equation}
  B_1^{\rm (in)}(r) = \frac{2a_1^{\rm (in)}}{r^2} \quad \mbox{and} \quad
  B_2^{\rm (in)}(r) = -\frac{a_{1,r}^{\rm (in)}}{r^2}\,.
\end{equation}
With the magnetic field determined both in the interior and exterior regions, the Faraday tensor for the background field 
becomes
\begin{equation}
 F_{\mu\nu}=\epsilon_{\mu\nu\alpha\beta} B^{\alpha}u^{\beta} =r^2 \sin \theta  \left(
 \begin{array}{cccc}
 0  &  0              &  0              &    0           \\
 0  &  0              &  0              &  B_2 \sin\theta\\
 0  &  0              &  0              & -B_1 \cos\theta\\
 0  & -B_2 \sin\theta &  B_1 \cos\theta &    0           \\
 \end{array}
 \right)\,.
 \label{eq:Fmn}
 \end{equation}

\section{Perturbation Equations}
\label{sec:Perturbation}

We consider small perturbations of both the gravitational and electromagnetic fields, which can be described as
\begin{eqnarray}
{\tilde g}_{\mu\nu}&=& g_{\mu\nu}+h_{\mu\nu}, \\
{\tilde F}_{\mu\nu}&=& F_{\mu\nu}+f_{\mu\nu},
\end{eqnarray}
where $g_{\mu\nu}$ and $F_{\mu\nu}$ are the background quantities derived in the previous section. The tensors
 $h_{\mu\nu}$ and $f_{\mu\nu}$ denote small perturbations, i.e. $h_{\mu\nu}=\delta g_{\mu\nu}$ and 
$f_{\mu\nu} = \delta F_{\mu\nu}$. Linearization of the Einstein-Maxwell equations yields
\begin{eqnarray}
  \delta G_{\mu \nu} &=& 8\pi \delta \left(T_{\mu \nu}+E_{\mu\nu}\right) \, , \label{dEin1}\\
  \delta \left(T^{\mu\nu}_{\ \ ;\nu} + E^{\mu\nu}_{\ \ ;\nu}\right) &=& 0 \, , \label{dEin2}\\
  \partial_{\nu}\left[(-g)^{1/2} f^{\mu\nu}\right] &=& 4\pi \delta\left[(-g)^{1/2} J^{\mu}\right]
      - \partial_{\nu}\left[F^{\mu\nu} \delta (-g)^{1/2}\right] \, , \label{eq:dMax1} \\
  f_{\mu \nu, \lambda}+f_{\lambda \mu, \nu} +f_{\nu \lambda, \mu} &=& 0 \, . \label{eq:dMax2}
\end{eqnarray}

From Eq.~(\ref{eq:dMax1}), we find that the electromagnetic perturbations are driven by the gravitational perturbations 
via the term containing $\delta (-g)^{1/2}$ in the right hand side. On the other hand, for simplicity, we omit the back reaction 
of the electromagnetic perturbations on the gravitational perturbations, i.e. we set 
$\delta E_{\mu\nu}=\delta(E^{\mu\nu}_{\ \ ;\nu})=0$ in Eqs.~(\ref{dEin1}) and~(\ref{dEin2}). This simplification is based 
on the assumption that the energy stored in gravitational perturbations is considerably larger than that in electromagnetic 
perturbations, which are typically driven by the former. On the other hand, in the giant flares of SGR 1806-20 and SGR 
1900+14~\cite{Kouveliotou1998,Hurley1999,Israel2005}, whose peak luminosities are in the range of $10^{44}-10^{46}$ 
ergs s$^{-1}$, the dramatic rearrangement of the magnetic field might lead to emission of gravitational waves. 
Nevertheless, recent non-linear MHD simulations \cite{Lasky2011,Ciolfi2011,Zink2012,Lasky2012,Ciolfi2012} do not support 
these expectations.

The first two perturbative equations, Eq.~(\ref{dEin1}) and Eq.~(\ref{dEin2}), have been studied extensively in the past, 
in the absence of magnetic fields, both for stellar and black hole backgrounds (see, e.g.~\cite{Regge1957, Thorne1967, Zerilli1970, Kokkotas1992, Kojima1992,Allen1998}). 
Thus, in this article we use the perturbation equations derived in earlier works, and we derive the analytic form of 
the perturbation equations for the electromagnetic field together with their coupling to the gravitational perturbations.

The metric perturbations $h_{\mu\nu}$, in the Regge-Wheeler gauge \cite{Regge1957},
can be decomposed into tensor spherical harmonics in the following way
\begin{equation}
 h_{\mu\nu} =
 \sum_{l=2}^{\infty} \sum_{m=-l}^{l}\left(
 \begin{array}{cccc}
 e^{\nu} H_{0,lm}  &  H_{1,lm} & - h_{0,lm} {\sin^{-1}\theta}
 \partial_{\phi}
 & h_{0,lm} \sin\theta \, \partial_{\theta} \\
 \ast & e^{\lambda} H_{2,lm} & -h_{1,lm} {\sin^{-1}\theta} \partial_{\phi}
 & h_{1,lm} \sin\theta \, \partial_{\theta} \\
 \ast & \ast & r^2 K_{lm} & 0 \\
 \ast & \ast & 0 & r^2 \sin^2\theta K_{lm} \\
 \end{array}
 \right) Y_{lm}\,,
 \end{equation}
where $H_{0,lm}$, $H_{1,lm}$, $H_{2,lm}$ and $K_{lm}$ are the functions of $(t,r)$ describing the {\em polar perturbations}, while $h_{0,lm}$ and $h_{1,lm}$ describe the {\em axial} ones. On the other hand, the tensor harmonic expansion of the electromagnetic perturbations, $f_{\mu\nu}$, for the {\em Magnetic multipoles} (or {\em axial parity}) are given by
\begin{equation}
 f_{\mu\nu}^{\rm (M)} =
 \sum_{l=2}^{\infty} \sum_{m=-l}^{l}\left(
 \begin{array}{ccccccc}
 0 & & 0  & & f^{\rm (M)}_{02,lm} {\sin^{-1}\theta} \partial_{\phi}
 & &-f^{\rm (M)}_{02,lm} \sin\theta \, \partial_{\theta} \\
 0& & 0 & &f^{\rm (M)}_{12,lm} {\sin^{-1}\theta} \partial_{\phi}
 & &-f^{\rm (M)}_{12,lm} \sin\theta \, \partial_{\theta} \\
 \ast& &  \ast& & 0 && f^{\rm (M)}_{23,lm}\sin \theta \\
 \ast& & \ast && \ast && 0 \\
 \end{array}
 \right) Y_{lm}\,,
 \end{equation}
while the expansion for the {\em Electric multipoles} (or {\em polar parity}) can be written as
\begin{equation}
 f_{\mu\nu}^{\rm (E)} =
 \sum_{l=2}^{\infty} \sum_{m=-l}^{l}\left(
 \begin{array}{ccccccc}
 0    & & f^{\rm(E)}_{01,lm}  && f^{\rm (E)}_{02,lm} \partial_{\theta} && f^{\rm (E)}_{02,lm} \partial_{\phi} \\
 \ast &&  0        &  & f^{\rm (E)}_{12,lm} \partial_{\theta}& &f^{\rm (E)}_{12,lm} \partial_{\phi} \\
 \ast &&  \ast      & & 0                         &  & 0 \\
 \ast &&  \ast      & & 0                         &  & 0 \\
 \end{array}
 \right) Y_{lm}\,.
 \end{equation}
Hereafter, the quantities describing the magnetic- and electric-type electromagnetic perturbations will be 
denoted with the indices (M) and (E), respectively. We point out that $h_{\mu\nu}$ is a symmetric tensor, 
while both $f_{\mu\nu}^{\rm (M)}$ and $f_{\mu\nu}^{\rm (E)}$ are anti-symmetric tensors, i.e. $f_{\mu\nu}^{\rm (M)}= 
-f_{\nu\mu}^{\rm (M)}$ and $f_{\mu\nu}^{\rm (E)}= -f_{\nu\mu}^{\rm (E)}$. From the perturbed Maxwell 
equations, Eqs.~(\ref{eq:dMax2}), we can obtain the following relations connecting the above perturbative functions:
\begin{gather}
 f^{\rm(M)}_{12,lm} = \frac{1}{\Lambda} \frac{\partial {f}^{\rm(M)}_{23,lm}}{\partial r} \quad \mbox{and} \quad
     {f}^{\rm(M)}_{02,lm}= \frac{1}{\Lambda} \frac{\partial {f}^{\rm(M)}_{23,lm}}{\partial t}\,, \label{magn_2} \\
 f^{\rm(E)}_{01,lm}= \frac{\partial {f}^{\rm (E)}_{02,lm}} {\partial r}
     - \frac{\partial {f}^{\rm (E)}_{12,lm}} {\partial t}\,,    \label{evol_f12E_1}
\end{gather}
where $\Lambda \equiv l(l+1)$. Notice that $f^{\rm(M)}_{23,lm}$ and ${\tilde \Psi}$, defined as 
\begin{equation}
 {\tilde \Psi} = -\frac{r^2}{\Lambda}{f}^{\rm(E)}_{01,lm} \,, \label{eq:gauge_f01E}
\end{equation}
are {\em gauge invariant} variables (see Eq. (II-27) in Ref.~\cite{CPM-I} and Eq. (II-11) in Ref.~\cite{CPM-II}).

\subsection{Perturbations of a Dipole  Magnetic Field: Exterior region}

In the exterior vacuum region, we adopt the condition 
$\delta J^{\mu} =0$. With this condition, the perturbed electromagnetic fields will be determined via the linearized form of  
Maxwell's equations, Eqs.~(\ref{eq:dMax1}), (assuming that $J^{\mu} = \delta J^{\mu} = 0$) 
\begin{equation}
  \partial_{\nu}\left[(-g)^{1/2} f^{\mu\nu}\right] = 
      - \frac{1}{2}\partial_{\nu}\left[(-g)^{1/2} F^{\mu\nu} g^{\alpha\beta}h_{\alpha\beta}\right]\,,
      \label{eq:dMaxEx}
\end{equation}
together with the perturbed Maxwell equation (\ref{eq:dMax2}). Equation~(\ref{eq:dMaxEx}) for  $\mu=t$ and $\mu=r$ can be written down as
\begin{eqnarray}
  \sum_{l,m}\left\{ A^{(I,\rm E)}_{lm} Y_{lm} + {\tilde A}^{(I,A)}_{lm} \cos \theta Y_{lm}
      + B^{(I,A)}_{lm} \sin \theta \partial_\theta Y_{lm}
      + C^{(I,P)}_{lm} \partial_\phi Y_{lm} \right\}=0 \ \ \ (I = 0,1)\,, \label{pert_0}
\end{eqnarray}
where the indices ``$A$'' and ``$P$'' stand for {\em axial} and {\em polar} gravitational perturbative quantities, and 
obviously ``$I$'' stands for the $t$ and $r$ components of Eq. (\ref{eq:dMaxEx}). The coefficients of Eq. (\ref{pert_0}) have 
the following expressions
\begin{eqnarray}
 A^{(0,\rm E)}_{lm} &=& \frac{1}{2}\left(\nu'+\lambda'-\frac{4}{r}\right)f_{01,lm}^{\rm (E)}
     - {f_{01,lm}^{\rm (E)}}' + \frac{\Lambda}{r^2}e^{\lambda}f_{02,lm}^{\rm (E)}\,,  \\
 {\tilde A}^{(0,A)}_{lm} &=& \frac{\Lambda}{r^2}e^{\lambda} B_1 h_{0,lm}\,,  \\
 B^{(0,A)}_{lm}          &=& \left[-\frac{1}{2}\left(\nu'+\lambda'-\frac{4}{r}\right)B_2 + {B_2}'
     + \frac{1}{r^2}e^{\lambda}B_1 \right]h_{0,lm} + B_2 {h_{0,lm}'}\,,  \\
 C^{(0,P)}_{lm}          &=& -B_2 H_{1,lm}\,, \\
 A^{(1,\rm E)}_{lm}          &=& r^2 \dot{f}_{01,lm}^{\rm (E)} - \Lambda e^{\nu}f_{12,lm}^{\rm (E)}\,,
     \label {evol_f01Es}  \\
 {\tilde A}^{(1,A)}_{lm} &=& -\Lambda e^{\nu} B_1 h_{1,lm}\,,  \\
 B^{(1,A)}_{lm}          &=& - e^{\nu}B_1 h_{1,lm} - r^2 B_2 \dot{h}_{0,lm}\,,  \\
 C^{(1,P)}_{lm}          &=& r^2 e^{\nu} B_2 H_{0,lm}\,.
\end{eqnarray}
One can decompose the equations above for a specific mode with fixed harmonic numbers $(l, m)$, by multiplying 
with $Y^*_{lm}$ and integrating over the two-sphere, i.e.
\begin{eqnarray}
 A^{(I,E)}_{lm}  + \I m C^{(I,P)}_{lm}
      +Q_{l m}\left[{\tilde A}^{(I,A)}_{l-1 m} + (l-1)B^{(I,A)}_{l-1 m}\right]
      +Q_{l+1 m}\left[{\tilde A}^{(I,A)}_{l+1 m} - (l+2)B^{(I,A)}_{l+1 m}\right]=0 \ \ \ (I = 0,1) \, .
\label{pert_1}
\end{eqnarray}
In a similar way, from the two remaining equations, i.e. Eq.~(\ref{eq:dMaxEx}) for $\mu=\theta$ and $\mu=\phi$, one gets the relations 
\begin{gather}
 \sum_{l,m} \left\{\left( \alpha_{lm} +{\tilde \alpha}_{lm}\cos \theta \right)\partial_\theta Y_{lm}
     - \left(\beta_{lm}+{\tilde \beta}_{lm} \cos \theta \right) (\partial_\phi Y_{lm}/\sin\theta) 
     + \eta_{lm}\sin\theta Y_{lm} + \chi_{lm}\sin \theta W_{lm} \right\} =0, \\
 \sum_{l,m} \left\{\left( \beta_{lm} +{\tilde \beta}_{lm}\cos \theta \right)\partial_\theta Y_{lm}
     + \left(\alpha_{lm}+{\tilde \alpha}_{lm} \cos \theta \right) (\partial_\phi Y_{lm}/\sin\theta) 
     + \zeta_{lm}\sin\theta Y_{lm} + \chi_{lm}\sin \theta X_{lm} \right\}=0,
\end{gather}
where
\begin{align}
 W_{lm} &= \left(\partial_{\theta}^2 - \cot\theta\partial_{\theta}
     - \frac{1}{\sin^2\theta}\partial_{\phi}\right) Y_{lm}, \\
 X_{lm} &= 2\partial_{\phi}\left(\partial_{\theta} - \cot\theta\right)Y_{lm} \, .
\end{align}
These equations lead to an extra set of evolution equations for a specific mode $(l,m)$ by multiplying 
with  $Y^*_{lm}$ and integrating over the two-sphere:
\begin{align}
 \Lambda \alpha_{lm} &- im \left[ {\tilde \beta}_{l m}+\zeta_{l m}\right] \nonumber \\
     &+ Q_{lm}(l+1)\left[(l-2)(l-1)\chi_{l-1 m}
                           +(l-1){\tilde \alpha}_{l-1 m} -\eta_{l-1 m} \right] \nonumber \\
     &- Q_{l+1 m} l \left[(l+2)(l+3)\chi_{l+1 m}
                           -(l+2){\tilde \alpha}_{l+1m} -\eta_{l+1 m} \right] = 0, \label{pert_2} \\
 \Lambda \beta_{lm} &+ im \left[ (l-1)(l+2)\chi_{l m}+{\tilde \alpha}_{l m}+\eta_{l m}\right] \nonumber \\
     &+ Q_{lm}(l+1)\left[(l-1){\tilde \beta}_{l-1 m}-\zeta_{l-1 m} \right]
      + Q_{l+1 m} l \left[(l+2){\tilde \beta}_{l+1 m} +\zeta_{l+1 m} \right] = 0, \label{pert_3}
\end{align}
where the coefficients are given by
\begin{align}
 \alpha_{lm}&= \frac{1}{2}\left(\lambda' - \nu'\right)f_{12,lm}^{\rm (E)} + e^{\lambda-\nu}\dot{f}_{02,lm}^{\rm (E)}
     - {f_{12,lm}^{\rm (E)}}'\,, \\
 \beta_{lm} &= \frac{1}{2}\left(\nu' - \lambda'\right)f_{12,lm}^{\rm (M)} - e^{\lambda-\nu}\dot{f}_{02,lm}^{\rm (M)}
     + {f_{12,lm}^{\rm (M)}}' - \frac{1}{r^2}e^{\lambda}f_{23,lm}^{\rm (M)}\,, \label{eq:beta_lm}  \\
 {\tilde \alpha}_{lm} &= \left[\frac{1}{2}\left(\lambda' - \nu'\right)B_1 - {B_1}' + B_2\right]h_{1,lm}
     - B_1 h'_{1,lm} + e^{\lambda-\nu}B_1 \dot{h}_{0,lm}\,, \\
 {\tilde \beta}_{lm} &= e^{\lambda} B_1 K_{lm},  \label{eq:tbeta_lm} \\
 \eta_{lm}  &= \frac{\Lambda}{2} B_2 h_{1,lm}\,, \\
 \chi_{lm}  &= \frac{1}{2} B_2 h_{1,lm}\,,  \\
 \zeta_{lm} &= \left[\frac{r^2}{2}\left(\lambda' - \nu'\right)B_2 - 2rB_2 - r^2 {B_2}'\right]H_{0,lm}
     - r^2B_2 H'_{0,lm} - e^{\lambda}B_1 K_{lm} + e^{-\nu} r^2 B_2 \dot{H}_{1,lm}\,. \label{eq:zeta_lm}
\end{align}

\subsection{Perturbations of a Dipole  Magnetic Field: Interior region}

In the stellar interior, because we have adopted the ideal MHD approximation for which $F_{\mu\nu}u^{\nu}=0$, the components of 
the perturbed electromagnetic field tensor are determined by using the perturbed Maxwell equation (\ref{eq:dMax2}), i.e.
\begin{equation}
  f_{0\mu} = e^{\nu/2} F_{\mu\nu} \delta u^{\nu}\,, \label{eq:in-Max}
\end{equation}
where $\delta u^{\mu}$ is the perturbed fluid  4-velocity, defined as
\begin{equation}
 \delta u^{\mu} = \left(\frac{1}{2}e^{-\nu/2}H_{0,lm},R_{lm},
     V_{lm}\partial_{\theta}-U_{lm}\sin^{-1}\theta\partial_{\phi},
     V_{lm}\sin^{-2}\theta\partial_{\phi} + U_{lm}\sin^{-1}\theta\partial_{\theta}\right) Y_{lm}\,.
\end{equation}
From Eq.~(\ref{eq:in-Max}) one can get the following equations
\begin{eqnarray}
 && \sum_{l,m} \left\{f_{01,lm}^{\rm (E)} Y_{lm} - r^2 B_2 e^{\nu/2} \left(V_{lm}\partial_{\phi}Y_{lm}
     + U_{lm} \sin\theta \partial_{\theta}Y_{lm}\right)\right\} = 0\,, \label{eq:int-1}\\
 && \sum_{l,m} \left\{\left({\cal A}_{lm} + \tilde{\cal A}_{lm} \cos\theta\right)\partial_{\theta} Y_{lm}
     - \left({\cal B}_{lm} + \tilde{\cal B}_{lm} \cos\theta \right)(\partial_{\phi}Y_{lm}/\sin\theta)\right\} = 0\, , \label{eq:int-2} \\
 && \sum_{l,m} \left\{\left({\cal B}_{lm} + \tilde{\cal B}_{lm} \cos\theta \right) \partial_{\theta} Y_{lm}
     + \left({\cal A}_{lm} + \tilde{\cal A}_{lm} \cos\theta\right)(\partial_{\phi} Y_{lm}/\sin\theta)
     + \tilde{\cal C}_{lm} (\sin\theta Y_{lm})\right\} = 0\, , \label{eq:int-3}
\end{eqnarray}
where the coefficients ${\cal A}_{lm}$ and ${\cal B}_{lm}$ are functions of the perturbed electromagnetic fields, 
while $ \tilde{\cal A}_{lm}$, $ \tilde{\cal B}_{lm},$ and $ \tilde{\cal C}_{lm}$ are functions of the perturbed matter fluid 4-velocity.
The expressions for these coefficients are
\begin{eqnarray}
 {\cal A}_{lm}       &=& f_{02,lm}^{\rm (E)}\,, \\
 {\cal B}_{lm}       &=& -f_{02,lm}^{\rm (M)}\,, \\
 \tilde{\cal A}_{lm} &=& r^2 B_1 e^{\nu/2} U_{lm}\,, \\
 \tilde{\cal B}_{lm} &=& -r^2 B_1 e^{\nu/2} V_{lm}\,, \\
 \tilde{\cal C}_{lm} &=& r^2 B_2 e^{\nu/2} R_{lm}\,.
\end{eqnarray}
By multiplying Eqs. (\ref{eq:int-1}),  (\ref{eq:int-2}), and (\ref{eq:int-3}) with $Y_{lm}^{*}$ and integrating over the two-sphere 
we can obtain the following system of equations that depends only on $r$ 
\begin{gather}
  f_{01,lm}^{\rm (E)} - r^2 B_2 e^{\nu/2} \left[im V_{lm} + Q_{lm} (l-1) U_{l-1m} - Q_{l+1m} (l+2) U_{l+1m}\right] = 0\, ,
      \label{eq:in_mag1} \\
  \Lambda {\cal A}_{lm} - im[\tilde{\cal B}_{lm} + \tilde{\cal C}_{lm}] + Q_{lm}(l-1)(l+1)\tilde{\cal A}_{l-1m}
     + Q_{l+1m}l(l+2)\tilde{\cal A}_{l+1m} = 0\, , \label{eq:in_mag2} \\
  \Lambda {\cal B}_{lm} + im\tilde{\cal A}_{lm} + Q_{lm}(l+1)[(l-1)\tilde{\cal B}_{l-1m} - \tilde{\cal C}_{l-1m}]
     + Q_{l+1m}l[(l+2)\tilde{\cal B}_{l+1m} + \tilde{\cal C}_{l+1m}] = 0\, , \label{eq:in_mag3}
\end{gather}
where
\begin{equation}
 Q_{lm} \equiv \sqrt{\frac{(l-m)(l+m)}{(2l-1)(2l+1)}} \, .
\end{equation}

Finally, we should compute Eqs.~(\ref{pert_1}), (\ref{pert_2}), and~(\ref{pert_3}) for the exterior region, and Eqs.~(\ref{eq:in_mag1}), (\ref{eq:in_mag2}), and~(\ref{eq:in_mag3}) for the interior region of 
the star. From this system of equations, we can see the specific couplings between the electromagnetic and gravitational perturbations. 
For example, an electromagnetic perturbation of specific parity with harmonic indices $(l,m)$ depends on the gravitational perturbations of the same parity with $(l,m)$ as well as the gravitational perturbations of the opposite parity with $(l\pm1,m)$. In other words, for the special and simpler case of  axisymmetric perturbations ($m=0$), we arrive at the following conclusions:
1) Dipole electric (polar) electromagnetic perturbations  will be driven by axial quadrupole gravitational perturbations, and 
2) Dipole magnetic (axial) electromagnetic perturbations will be driven by polar quadrupole and radial gravitational perturbations.
These two types of couplings will be discussed in detail in the next sections.

\subsection{Junction conditions for perturbed electro-magnetic fields}

In order to close the system of equations derived in the previous subsection, we should impose appropriate junction conditions on 
the stellar surface. Such junction conditions for the perturbed electromagnetic fields can be derived from the conditions 
\begin{eqnarray}
 n^{\mu}\delta B_{\mu}^{({\rm in})} &=& n^{\mu} \delta B_{\mu}^{({\rm ex})}\,, \\
 q_{\mu}^{\ \nu} \delta E_{\nu}^{({\rm in})} &=& q_{\mu}^{\ \nu} \delta E_{\nu}^{({\rm ex})}\,,
\end{eqnarray}
where $n^{\mu}$ is the unit outward normal vector to the stellar surface, while $q_{\mu}^{\ \nu}$ is the corresponding projection 
tensor associated with $n^{\mu}$. These junction conditions lead to the following set of equations: 
\begin{eqnarray}
 f_{23}^{{\rm (M)}({\rm in})} &=& f_{23}^{{\rm (M)}({\rm ex})}\,, \label{eq:junction1} \\
 f_{02}^{{\rm (M)}({\rm in})} &=& f_{02}^{{\rm (M)}({\rm ex})} = 0\,, \label{eq:junction2} \\
 f_{02}^{{\rm (E)}({\rm ex})} &=& 0\,. \label{eq:junction3} 
\end{eqnarray}

\section{Dipole Perturbations of a Magnetic Field on a Stellar Background}
\label{sec:Pert_Star}

In the previous section, we provided the general form of the perturbative equations.
In order to focus on a simple case, we only consider axisymmetric perturbations ($m=0$) in this section. In this way, the various couplings become less complicated. 
Under these conditions, we study the excitation of dipole electric perturbations driven by axial gravitational ones and dipole magnetic perturbations 
driven by polar gravitational ones. These perturbative modes are actually the most important ones from the energetic point of view.

\subsection{Dipole Electric Perturbations driven by Axial Gravitational Perturbations} 
\label{sec:Pert_Star_1}

Here, we consider only dipole ``electric type" perturbations driven by quadrupole axial gravitational perturbations. 
Since we neglect the back reaction of electromagnetic perturbations on the gravitational ones, the quadrupole axial 
gravitational perturbations of a spherically symmetric star can be described by a single wave equation \cite{Thorne1967,CF1991}, 
which is given by
\begin{equation}
 \frac{\partial^2 X_{lm}}{\partial t^2}-\frac{\partial^2
    X_{lm}}{\partial r_*^2}+e^{\nu}\left(\frac{\Lambda}{r^2}-\frac{6m}{r^3}+4\pi(\rho-p)\right)
    X_{lm}=0\,,  \label{eq:evol_X}
\end{equation}
where
\begin{equation}
 X_{lm}=\frac{e^{(\nu-\lambda)/2}}{r}h_{1,lm} \quad \mbox{and} \quad
    \frac{\partial }{\partial r}= e^{(\lambda-\nu)/2}\frac{\partial }{\partial r_*}\,.
    \label{eq:evol_h0}
\end{equation}
Note that $r_*$ is the tortoise coordinate defined as $r_* = r+2M\ln(r/2M-1)$. Since there are no fluid oscillations if the matter is 
assumed to be described as a perfect fluid (unless we introduce rotation), the spacetime only contains pure spacetime modes, i.e. the so-called 
$w$-modes~\cite{Kokkotas1992,CF1991,Kokkotas1994}. In this case, the axial component of the fluid perturbation, $U_{lm}$, has the form
$U_{lm} = -e^{-\nu/2}h_{0, lm}/r^2$, while the component of $h_{0,lm}$ is computed from the equation
\begin{eqnarray}
\frac{\partial}{\partial t} h_{0,lm} & = &
    e^{(\nu-\lambda)/2}X_{lm}+ r\frac{\partial}{\partial r_*} X_{lm},
    \label{evol_h0}
\end{eqnarray}
which is used later to simplify the coupling terms between the two types of perturbations.

On the other hand, in the same way as in the case of electromagnetic perturbations in the exterior region, Eqs. (\ref{evol_f12E_1}) and (\ref{pert_1}) for $I=1$, 
and Eq.~(\ref{pert_2}) lead to three simple evolution equations for the three perturbation functions  $f_{12,10}^{\rm (E)}$, $f_{01,10}^{\rm (E)}$, and $f_{02,10}^{\rm (E)}$:
\begin{align}
 \frac{\partial f^{\rm (E)}_{12,10}}{\partial t} &= e^{-\nu} \frac{\partial f_{02,10}^{\rm (E)}}{\partial r_*}
     - f_{01,10}^{\rm (E)}\,, \\
 \frac{\partial f_{01,10}^{\rm (E)}}{\partial t} &= \frac{2}{r^2}e^{\nu}f_{12,01}^{\rm (E)} + S_{20}^{(1)}\,, \\
 \frac{\partial f_{02,10}^{\rm (E)}}{\partial t} &= e^{\nu} \frac{\partial f_{12,10}^{\rm (E)}}{\partial r_*}
     + \nu' e^{2\nu} f_{12,10}^{\rm (E)} + S_{20}^{(2)}\,,
\end{align}
where $S_{20}^{(1)}$ and $S_{20}^{(2)}$ are the source terms describing the coupling of the electromagnetic perturbations 
with the gravitational ones, and are given by
\begin{align}
 S_{20}^{(1)} &= 3Q_{20}\left[\left(\frac{1}{r}B_1 - e^{\nu} B_2 \right)X_{20}
     - rB_2\frac{\partial X_{20}}{\partial r_*}\right]\,, \label{eq:S1-axial} \\
 S_{20}^{(2)} &= \frac{3}{2}Q_{20} re^{\nu}{B_1}'X_{20}\,. \label{eq:S2-axial}
\end{align}
In order to derive second-order wave-type equations for the electromagnetic perturbations, we introduce a new function: 
$\Psi_{lm}=\Psi_{lm}(t,r)$, given by
\begin{equation}
  \Psi_{lm} = e^{\nu} f_{12,lm}^{\rm (E)}\,.
\end{equation}
With this variable, the above evolution equations can be written as
\begin{align}
 \frac{\partial \Psi_{10}}{\partial t} &= \frac{\partial f_{02,10}^{\rm (E)}}{\partial r_*}
     -e^{\nu}f_{01,lm}^{\rm (E)}\,, \\
 \frac{\partial f_{01,10}^{\rm (E)}}{\partial t} &= \frac{2}{r^2}\Psi_{10} + S_{20}^{(1)} \label{evol_f01_ns}\,, \\
 \frac{\partial f_{02,10}^{\rm (E)}}{\partial t} &= \frac{\partial \Psi_{10}}{\partial r_*} + S_{20}^{(2)}\,.
\end{align}
From this system of evolution equations, one can construct a single wave-type equation for the ``electric'' perturbations
\begin{eqnarray}
 \frac{\partial^2 \Psi_{10}}{\partial t^2} - \frac{\partial^2 \Psi_{10}}{\partial r_*^2}
     + \frac{2}{r^2}e^{\nu}\Psi_{10} =S_{20}^{\rm (E)}\,, \label{eq:wave-mp-Psi}
\end{eqnarray}
where the source term $S_{20}^{\rm (E)}$ is given by
\begin{eqnarray}
 S_{20}^{\rm (E)} = \frac{\partial S_{20}^{(2)}}{\partial r_*} - e^{\nu}S_{20}^{(1)} \, . \label{eq:Se-axial}
\end{eqnarray}
Without the coupling term, this wave equation outside the star is the well-known Regge-Wheeler equation for 
electromagnetic perturbations. It should be pointed out that $\Psi$ is not a gauge-invariant quantity while the 
function ${\tilde \Psi}$ given by Eq.~(\ref{eq:gauge_f01E}) is a gauge invariant variable, where both variables 
$\Psi$ and ${\tilde \Psi}$ can be related to each other via the evolution equation (\ref{evol_f01_ns}), i.e.
\begin{equation}
 \frac{\partial {\tilde \Psi}_{10}}{\partial t} = -\Psi_{10} - \frac{r^2}{2}S_{20}^{(1)}\, .
\end{equation}

Finally, the electromagnetic perturbations in the interior region are determined from Eqs.~(\ref{eq:in_mag1}), 
(\ref{eq:in_mag2}), and (\ref{evol_f12E_1}), i.e.
\begin{align}
 f_{01,10}^{\rm (E)} &= B_2 S_{20}^{(3)} \, , \\
 f_{02,10}^{\rm (E)} &= \frac{1}{2} B_1 S_{20}^{(3)} \, , \\
 \frac{\partial f_{12,10}^{\rm (E)}}{\partial t} &= \frac{\partial f_{02,10}^{\rm (E)}}{\partial r}
     - f_{01,10}^{\rm (E)} \, ,
\end{align}
where
\begin{equation}
 S_{20}^{(3)} = -3Q_{20} r^2 e^{\nu/2} U_{20} \, .
\end{equation}

\subsection{Dipole Magnetic Perturbations driven by Polar Gravitational Perturbations}
\label{sec:Pert_Star_2}

As it was mentioned earlier in Sec.~\ref{sec:Perturbation} for the case of axisymmetric perturbations, the ``magnetic (axial) type" 
perturbations of the electromagnetic field with harmonic index $l$ are driven by polar gravitational perturbations with harmonic 
index $l \pm 1$. Here, we consider the axisymmetric perturbations ($m=0$) for the dipole ($l=1$) electromagnetic fields, which are 
driven by quadrupole ($l=2$) gravitational perturbations.

For the description of the perturbations of the spacetime and the stellar fluid, we adopt the formalism derived by Allen $et$ 
$al$. in~\cite{Allen1998}. In this formalism, the perturbations are described by three coupled wave-type equations, in such a way that 
two equations describe the perturbations of the spacetime and the other one the fluid perturbations. In addition to these three 
wave equations, there is also a constraint equation. The two wave-type equations for the spacetime variables are
\begin{align}
 -\frac{\partial ^2 S_{lm}}{\partial t^2} + \frac{\partial ^2 S_{lm}}{\partial r_*^2}
     + \frac{2 e^\nu}{r^3}\left[ 2 \pi r^3 (\rho+3p) + m - (n+1)r \right]S_{lm}
     = -\frac{4 e^{2\nu}}{r^5} \left[\frac{(m+4\pi p r^3)^2}{r-2m} + 4\pi \rho r^3 - 3m \right] F_{lm}\,,
     \label{eq:wave_ns_m-p-S}
\end{align}
\begin{align}
 -\frac{\partial ^2 F_{lm}}{\partial t^2} + \frac{\partial ^2 F_{lm}}{\partial r_*^2}
     +& \frac{2 e^\nu}{r^3} \left[2 \pi r^3 (3 \rho+ p) + m - (n+1)r \right] F_{lm} \nonumber \\
     =& - 2\left[ 4 \pi r^2 (p+\rho) - e^{-\lambda} \right] S_{lm}
     + 8 \pi (\rho+p) {r e^{\nu}} \left(1- \frac{1}{C_s^2} \right) H_{lm} \,, \label{eq:wave_ns_m-p-F}
\end{align}
where $F_{lm}$, $S_{lm}$, and $H_{lm}$ are given by
\begin{align}
 F_{lm}(t,r) &= rK_{lm}\,, \\
 S_{lm}(t,r) &= \frac{e^{\nu}}{r}\left(H_{0,lm} -K_{lm}\right)\,, \\
 H_{lm}(t,r) &=\frac{\delta p_{lm}}{\rho+p}\,,
\end{align}
while $\delta p_{lm}$ is the perturbation in the pressure, $n \equiv (l-1)(l+2)/2$, and $C_s$ is the sound speed. 
On the other hand, the wave equation for the perturbed relativistic enthalpy $H_{lm}$, describing the fluid perturbations, is 
\begin{eqnarray}
 - \frac{1}{C_s^2}\frac{\partial ^2 H_{lm}}{\partial t^2} + \frac{\partial ^2 H_{lm}}{\partial r_*^2}
     &+& \frac{e^{(\nu+\lambda)/2}}{r^2}\left[(m + 4\pi p r^3)\left(1-\frac{1}{C_s^2}\right) + 2 (r-2m) \right]
         \frac{\partial H_{lm}}{\partial r_*} \nonumber \\
     &+& \frac{2 e^\nu}{r^2} \left[ 2 \pi r^2 (\rho+p)\left(3 + \frac{1}{C_s^2} \right) - (n+1) \right] H_{lm}
         \nonumber \\
     &=& (m+4 \pi p r^3)\left(1-\frac{1}{C_s^2}\right) \frac{e^{(\lambda-\nu)/2}}{2 r}
         \left(\frac{e^\nu}{r^2}\frac{\partial F_{lm}}{\partial r_*} - \frac{\partial S_{lm}}{\partial r_*}
         \right) \nonumber \\
     &+& \left[\frac{(m+4\pi p r^3)^2}{r^2(r-2m)}\left(1+\frac{1}{C_s^2}\right) - \frac{ m+4\pi p r^3}{2 r^2} 
         \left(1-\frac{1}{C_s^2}\right)-4\pi r (3p+\rho) \right] S_{lm} \nonumber \\
     &+& \frac{e^\nu}{r^2}\left[\frac{2(m+4\pi p r^3)^2}{r^2(r-2m)}\frac{1}{C_s^2}
         - \frac{ m+4\pi p r^3}{2 r^2}\left(1-\frac{1}{C_s^2} \right) -4\pi r (3p+\rho)\right] F_{lm}\,.
 \label{eq:wave_ns_m-p-H}
\end{eqnarray}
This third wave equation (\ref{eq:wave_ns_m-p-H}) is valid only inside the star, while the first two are simplified considerably
outside the star, which can be reduced to a single wave-type equation, i.e. the Zerilli equation (see~\cite{Allen1998} 
and \S \ref{sec:Pert_BH2}).
Finally, the Hamiltonian constraint,
\begin{eqnarray}
 \frac{\partial ^2 F_{lm}}{\partial r_*^2}
     &-& \frac{e^{(\nu+\lambda)/2}}{r^2} \left( m + 4 \pi r^3 p \right) \frac{\partial F_{lm}}{\partial r_*}
         + \frac{e^\nu}{r^3} \left[ 12\pi r^3 \rho - m - 2(n+1)r \right] F_{lm}
         \nonumber \\
     &-& r e^{-(\nu+\lambda)/2}\frac{\partial S_{lm}}{\partial r_*} + \left[8\pi r^2(\rho+p) -(n+3)
         + \frac{4m}{r} \right] S_{lm} + \frac{8 \pi r}{C_s^2} e^\nu (\rho+p) H_{lm} = 0\,,
 \label{Hamilton}
\end{eqnarray}
can be used for setting up initial data and monitoring the evolution of the coupled system.

Regarding the quadrupole gravitational perturbations, the perturbation equation for the  ``magnetic type"  dipole
in the exterior region is obtained from Eq.~(\ref{pert_3}) as
\begin{equation}
 \frac{\partial^2 \Phi_{10}}{\partial t^2} - \frac{\partial^2 \Phi_{10}}{\partial r_*^2}
     + \frac{2}{r^2}e^{\nu}\Phi_{10} = S_{20}^{\rm (M)}\,, \label{eq:wave-magnetic}
\end{equation}
where $\Phi_{lm} \equiv f_{23,lm}^{\rm (M)}$ and
\begin{equation}
 S_{20}^{\rm (M)} = -Q_{20}e^{\nu}\left[\left(2B_2 + r {B_2}'\right)r^2S_{20}
     + \left(e^{\nu}B_2 + re^{\nu}{B_2}' - \frac{2}{r}B_1\right)F_{20}
     + rB_2 \frac{\partial F_{20}}{\partial r_*}\right]\,. \label{eq:source-magnetic}
\end{equation}
In order to derive the wave equation (\ref{eq:wave-magnetic}), we have used Eq.~(\ref{magn_2}) and the 
($r,\phi$)-component of the perturbed Einstein equations, i.e. $e^{-\nu}{\dot H_1} - H_0'+ K'- \nu' H_0 = 0$. 
We remark that the wave equation (\ref{eq:wave-magnetic}) without the source terms is the same as the one 
derived in~\cite{SYK2007,Sotani2009}. In addition, the other components of the electromagnetic perturbations, 
$f_{12,10}^{\rm (M)}$ and $f_{02,10}^{\rm (M)}$, can be determined with $\Phi_{10}$ via the relation (\ref{magn_2}).

Finally, from Eq.~(\ref{eq:in_mag3}) and  Eq.~(\ref{magn_2}), we can obtain the equation that determines the 
dipole ``magnetic type" perturbations for the interior region:
\begin{equation}
 \frac{\partial \Phi_{10}}{\partial t} = Q_{20}r^2e^{\nu/2}\left(B_2 R_{20} - 3B_1 V_{20}\right)\,, \label{eq:in-magnetic}
\end{equation}
where the perturbations of the fluid velocity, $R_{20}$ and $V_{20}$, in the source term are given by
\begin{align}
 \frac{\partial R_{20}}{\partial t} =& e^{\nu/2-\lambda} \left[\left(-\frac{11p + 3\rho}{2(p+\rho)}
     + \frac{3r\nu'}{2}\right)e^{-\nu}S_{20} - \frac{3}{2}re^{-\nu}S_{20}' 
     + \frac{3p - \rho}{2r^2(p+\rho)}\left(F_{20} - rF_{20}'\right) - H_{20}'\right]\,, \label{eq:R20}\\
 \frac{\partial V_{20}}{\partial t} =& \frac{1}{2r^2}e^{\nu/2}\left[re^{-\nu} S_{20}
     + \frac{\rho - 3p}{p+\rho}\frac{F_{20}}{r} - 2H_{20}\right]\,. \label{eq:V20}
\end{align}

\section{Perturbations of Dipole Magnetic Field on a BH background}
\label{sec:Pert_BH}

The perturbations of a dipole magnetic field on a Schwarzschild black hole background are described by the 
same set of perturbation equations as in the exterior region of the star except for the boundary conditions, i.e. 
the boundary conditions for the neutron star imposed on the stellar surface are Eqs. (\ref{eq:junction1}) -- (\ref{eq:junction3}), 
while for the black hole case one should impose the pure ingoing wave conditions at the event horizon. 
Then, even in the case of the black hole background, we observe the same coupling of the various harmonics of the 
electromagnetic and gravitational perturbations as for the neutron star background. That is, for the axisymmetric perturbations, 
the ``electric" dipole ($l=1$) perturbations of the electromagnetic fields will be driven by axial quadrupole ($l=2$) 
gravitational perturbations, while the ``magnetic'' dipole  ($l=1$) perturbations of the electromagnetic fields will be 
driven by polar quadrupole ($l=2$) gravitational ones. In this specific case, our study is similar to the work 
in~\cite{CMBD2004}, although they use a different formalism.

\subsection{Dipole Electric Perturbations driven by Axial Gravitational Perturbations (BH)}
\label{sec:Pert_BH1}

The axial quadrupole ($l=2$) gravitational perturbations are described by the Regge-Wheeler equation
\begin{equation}
 \frac{\partial^2 X_{lm}}{\partial t^2}-\frac{\partial^2 X_{lm}}{\partial r_*^2}
     + e^{\nu}\left(\frac{\Lambda}{r^2}-\frac{6M}{r^3}\right) X_{lm}=0\,,  \label{eq:wave_bh-e-a-X}
\end{equation}
where
\begin{equation}
 X_{lm}=\frac{e^{\nu}}{r}h_{1,lm}\,.
\end{equation}
In accordance with the results of Section \ref{sec:Pert_Star_1}, the perturbations of the electromagnetic 
fields will be described by a single wave equation, that is, the Regge-Wheeler equation for electromagnetic perturbations, 
give by
\begin{eqnarray}
 \frac{\partial^2 \Psi_{10}}{\partial t^2} - \frac{\partial^2 \Psi_{10}}{\partial r_*^2}
     + \frac{2}{r^2}e^{\nu}\Psi_{10} = S_{20}^{\rm (E)}\,,
\end{eqnarray}
where the source term becomes of the same form as in Section~\ref{sec:Pert_Star_1}: $\Psi_{lm} = e^{\nu} f_{12,lm}^{\rm (E)}$.

\subsection{Dipole Magnetic Perturbations driven by Polar Gravitational Perturbations (BH)}
\label{sec:Pert_BH2}

The equation describing the ``magnetic'' type perturbations driven by the gravitational perturbations 
is the same equation as the one derived for a neutron star background (see Eq.~(\ref{eq:wave-magnetic})), that is
\begin{eqnarray}
 \frac{\partial^2 \Phi_{10}}{\partial t^2} - \frac{\partial^2 \Phi_{10}}{\partial r_*^2}
     + \frac{2}{r^2}e^{\nu}\Phi_{10} = S_{20}^{\rm (M)}\,,
\end{eqnarray}
where $\Phi_{lm}=f_{23,lm}^{(M)}$, and the source term is also of the same form as in Eq.~(\ref{eq:source-magnetic}). 
The perturbative equation for the spacetime variables can be written in the form of the Zerilli equation
\begin{gather}
 \frac{\partial^2 Z_{lm}}{\partial t^2}-\frac{\partial^2 Z_{lm}}{\partial r_*^2} + V_Z(r)Z_{lm} = 0\,, \\
 V_{Z}(r) = \frac{2e^{\nu}\left[\Lambda_1^2(\Lambda_1 +1)r^3 + 3M\Lambda_1^2 r^2 + 9M^2\Lambda_1 r + 9M^3\right]}
     {r^3(r\Lambda_1 + 3M)^2},
\end{gather}
where $\Lambda_1\equiv (l+2)(l-1)/2$. Meanwhile, in the same way as for the neutron star background, one can also 
adopt $F_{lm}$ and $S_{lm}$ as the perturbation variables for the spacetime. In this case, the two wave equations 
simplify to become
\begin{align}
 \frac{\partial^2 S_{lm}}{\partial t^2} - \frac{\partial^2 S_{lm}}{\partial r_*^2} 
     + e^{\nu}\left(\frac{\Lambda}{r^2} - \frac{2M}{r^3}\right)S_{lm}
     &= -\frac{4M}{r^5}e^{\nu}\left(3-\frac{7M}{r}\right)F_{lm}\,, \\
 \frac{\partial^2 F_{lm}}{\partial t^2} - \frac{\partial^2 F_{lm}}{\partial r_*^2}
     + e^{\nu}\left(\frac{\Lambda}{r^2} - \frac{2M}{r^3}\right)F_{lm} &=-2e^{\nu}S_{lm}\,,
\end{align}
which have to be supplemented with the Hamiltonian constraint equation
\begin{eqnarray}
 \frac{\partial^2 F_{lm}}{\partial r_*^2}-\frac{M}{r^2}\frac{\partial F_{lm}}{\partial r_*}
    - \frac{\Lambda}{r^2}e^{\nu} F_{lm} -r\frac{\partial S_{lm}}{\partial r_*}
    - \frac{1}{2}\left(4e^{\nu}+\Lambda\right)S_{lm} = 0\,.
\end{eqnarray}
Note that there are useful relations between the perturbation variables ($F_{lm}$, $S_{lm}$) and the Zerilli function ($Z$), i.e.
\begin{eqnarray}
 F_{lm} &=& r\frac{dZ_{lm}}{dr_*}+\frac{\Lambda_1(\Lambda_1+1)r^2+3 \Lambda_1 Mr+6M^2}{r(\Lambda_1 r+3M)}Z_{lm}\,, \\
 S_{lm} &=& \frac{1}{r}\frac{dF_{lm}}{dr_*}-\frac{(\Lambda_1+2)r-M}{r^3}F_{lm}+\frac{(\Lambda_1+1)(\Lambda_1 r+3M)}{r^3}Z_{lm}\,,
\end{eqnarray}
which can be used in constructing initial data (since the Zerilli function is gauge invariant and unconstrained), or for the 
extraction of the Zerilli function.

\section{Applications}
\label{sec:Results}

As an application, we consider the case in which  dipole ``electric type" perturbations are driven by axial 
gravitational ones and present numerical results. First, we study the coupling on a Schwarzschild black hole background and 
later on the background of spherical neutron stars, as discussed in \S \ref{sec:Pert_BH1} and in \S \ref{sec:Pert_Star_1}, respectively. 
The more complicate cases that involve the driving of  ``magnetic type'' electromagnetic field perturbations driven by polar gravitational 
ones will be discussed elsewhere in the future.

\subsection{Perturbations on a Black-Hole Background}
\label{sec:Application1}

In order to calculate the waveforms in the black hole background, we need to modify the background magnetic field near the event horizon.
The reason for this is that the solution for a dipole magnetic field in vacuum diverges at the event horizon (see Eqs. (\ref{eq:B1-ex}) and (\ref{eq:B2-ex})). 
In fact, the isolated black hole cannot have magnetic fields due to the no hair theorem. But, according to the simulations of the accretion onto 
the black hole, the magnetic field can reach almost to the event horizon, because the accreting matter will fall into the black hole with 
infinite time \cite{TNM2011,MTB2012}. Thus, we adopt a simple modification of the dipole magnetic field near the event horizon, that is, we set 
$B_1(r)= B_1(6M)$ and $B_2(r)=B_2(6M)$ for $r\le 6M$, where the position at $r=6M$ corresponds to the innermost stable circular orbit for a 
test particle around the Schwarzschild black hole.
The magnetic dipole moment $\mu_d$ is identified with the normalized magnetic field strength $B_{15}$, defined as 
$B_{15}\equiv B_p/(10^{15} [G])$, where $B_p$ is the field strength at $r=6M$ and $\theta=0$.

We assume vanishing  electromagnetic perturbations i.e., $\Psi_{10}=\partial \Psi_{10}/\partial t=0$ at the initial time slice $t=0$, 
while the initial gravitational perturbations, $X_{20}$, are prescribed in terms of a Gaussian wave packet. Under these initial conditions, 
the electromagnetic waves will result from the coupling to the gravitational ones. In the numerical calculations, we adopt 
the iterated Crank-Nicholson method \cite{Teukolsky2000} with a grid choice of $\Delta r_* = 0.1M$ and $\Delta t = \Delta r_* / 2$ 
(see~\cite{Sotani2006} for the dependence of the choice of $\Delta r_*$ and $\Delta t$ on the waveforms).

The energy emitted in the form of either gravitational  ($E_{\rm GW}$) or  electromagnetic waves ($E_{\rm EM}$), is estimated by 
integrating the luminosity ($L_{{\rm GW},l}^{\rm (A)}$) for the axial gravitational waves and for electric type electromagnetic waves 
($L_{{\rm EM},l}^{\rm (E)}$), which are described by the following formulae \cite{CPM-I,CPM-II}
\begin{gather}
  L_{{\rm GW},l}^{\rm (A)} = \frac{1}{16\pi}\frac{(l-2)!}{(l+2)!}\left|\frac{\partial X_{l0}}{\partial t}\right|^2\,, \\
  L_{{\rm EM},l}^{\rm (E)} = \frac{1}{4\pi}\frac{(l+1)!}{(l-1)!}\left|\frac{\partial \Psi_{l0}}{\partial t}\right|^2 \, .
\end{gather}
In practice, we can find a relation between the energy emitted in gravitational and electromagnetic waves for a given initial 
spacetime perturbation, which has the form:
\begin{equation}
  E_{\rm EM}=\alpha {B_{15}}^{2} E_{\rm GW}\,, \label{eq:relation}
\end{equation}
where $\alpha$ is a ``proportionality constant''.

\begin{figure}[htbp]
\begin{center}
\begin{tabular}{cc}
\includegraphics[scale=0.5]{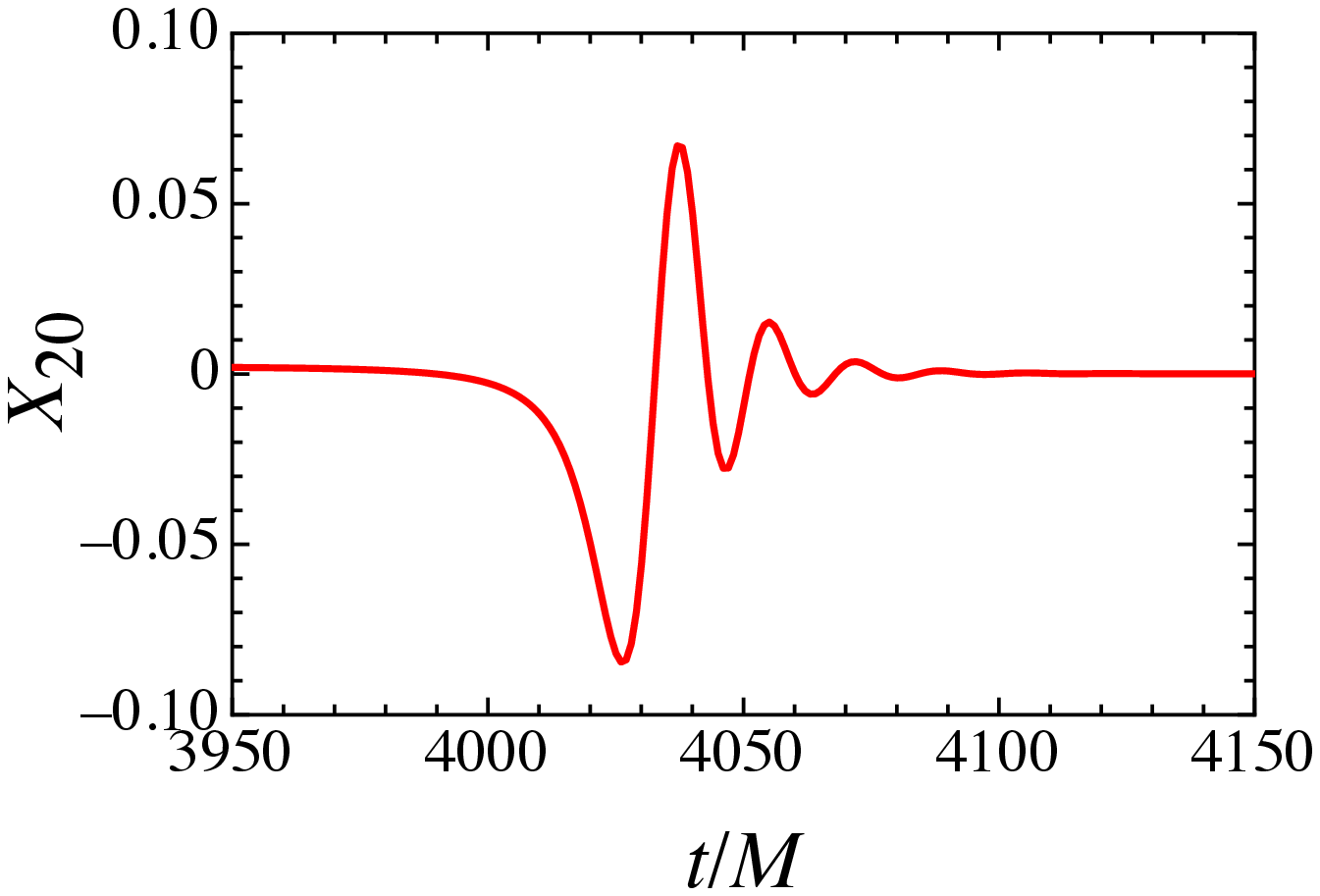} &
\includegraphics[scale=0.5]{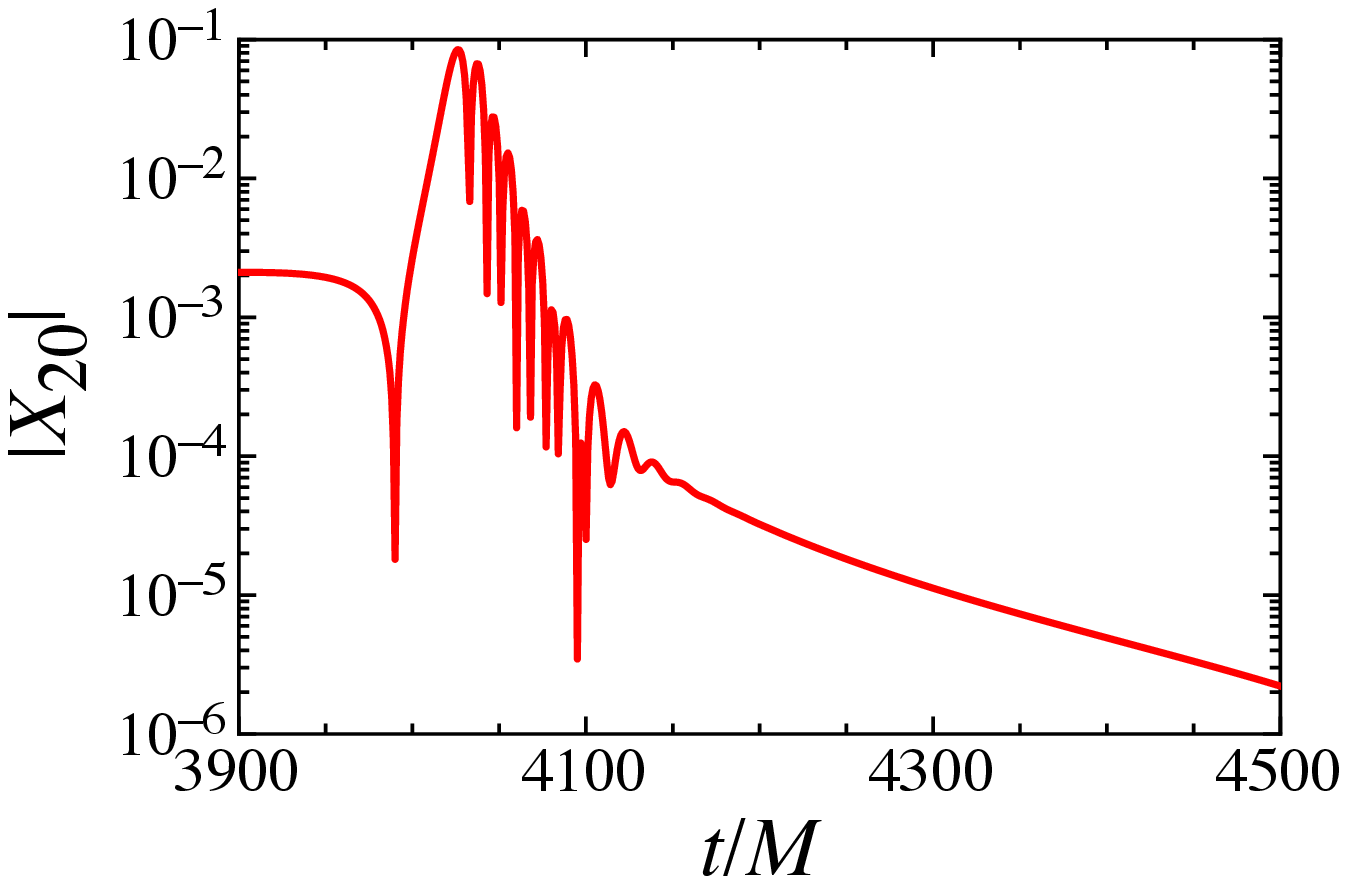} \\
\end{tabular}
\end{center}
\caption{
Gravitational waveform observed at $r=2000M$. In the right panel, we also show the absolute value
of $X_{20}$.
}
\label{fig:GW}
\end{figure}

\begin{figure}[htbp]
\begin{center}
\begin{tabular}{cc}
\includegraphics[scale=0.5]{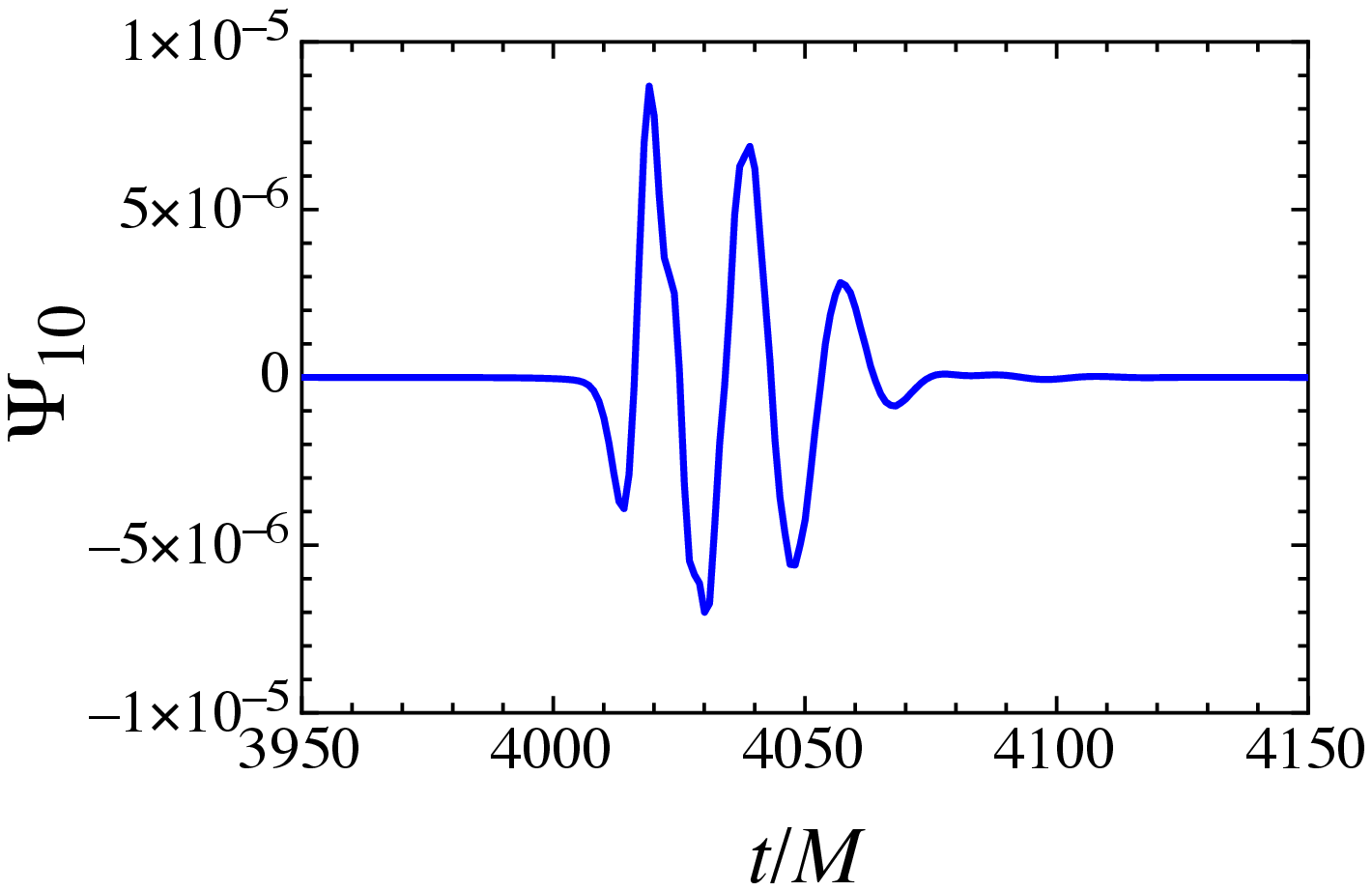} &
\includegraphics[scale=0.5]{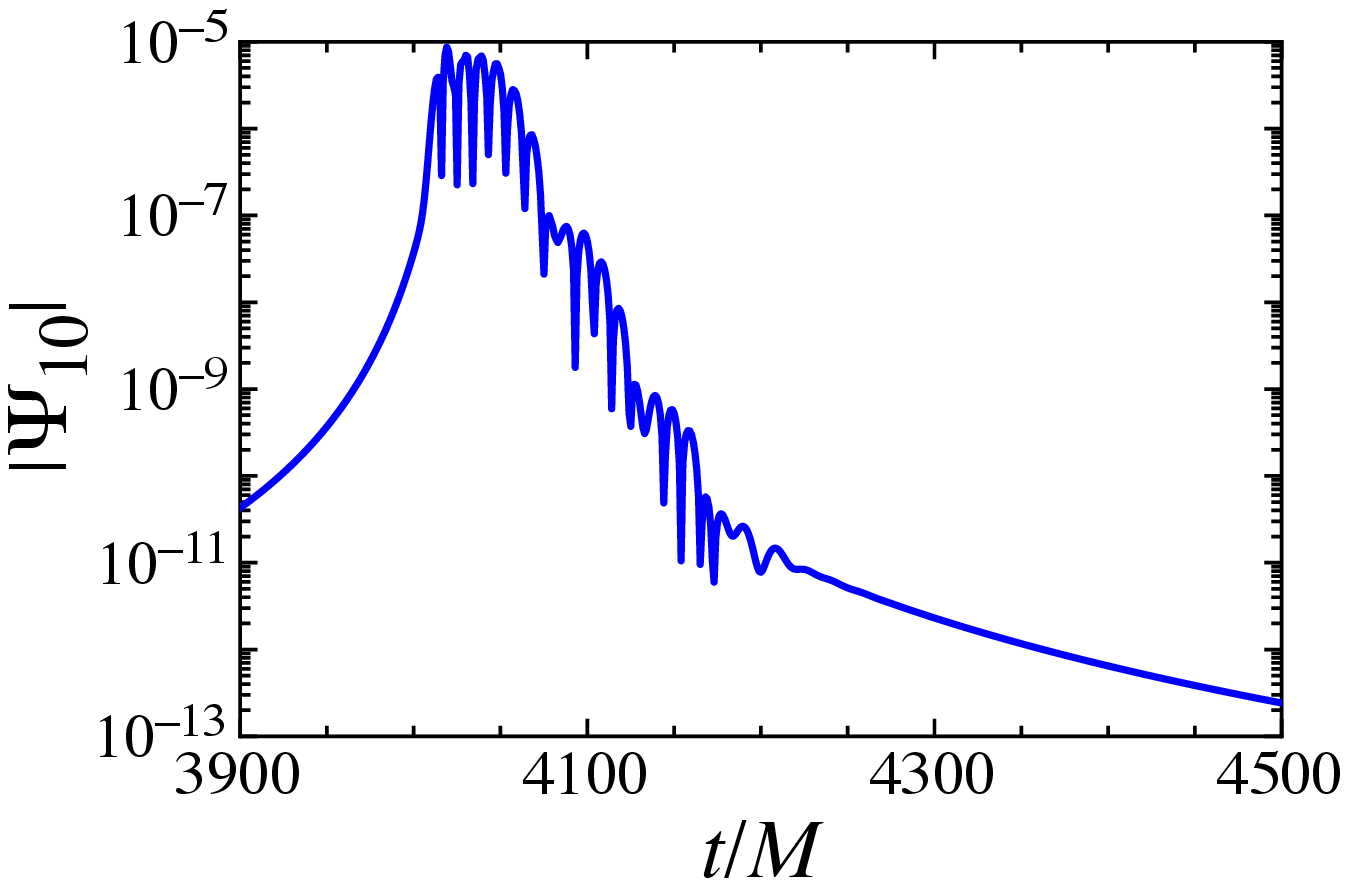} \\
\end{tabular}
\end{center}
\caption{
Waveform of the driven electromagnetic waves observed at $r=2000M$ for $B_p=10^{15}$ Gauss.
In the right panel, we also show the absolute value
of $\Psi_{10}$.
}
\label{fig:EMW-BH0}
\end{figure}

In the simulation that we describe we set the magnetic field strength to the value $B_p=10^{15}$ Gauss.  Fig. \ref{fig:GW} shows 
the waveform of the gravitational wave observed at $r=2000M$, the amplitude is normalized to correspond to an emitted energy of 
$E_{\rm GW}\approx 1.8\times 10^{49} \left(M/{50 M_\odot}\right)$ ergs. On the other hand, the waveforms of electromagnetic waves driven by the gravitational 
waves are shown in Fig. \ref{fig:EMW-BH0}. From this figure, we can observe somewhat complicated waveforms of electromagnetic waves 
due to the coupling with the gravitational waves. From the specific waveforms, one can estimate the value of the proportionality constant 
in the relation (Eq. (\ref{eq:relation})) to be $\alpha = 8.02\times 10^{-6}$. This efficiency might not be very high, but the 
radiated energy of gravitational waves can reach $\sim 10^{51}$ ergs for a black hole formation due to the merger of a neutron stars binary (see, e.g.~\cite{STU2003}).
In this case the strength of the magnetic field can be amplified by the Kelvin-Helmholz instability to reach values of  the order of $10^{15-17}$ Gauss  \cite{PR2006}.
Although this is not an ideal situation for the black hole case we are considering in this paper, if one adopts the above efficiency for the case of a black hole 
formed after merger, one can expect that  
energies of the order  of $\sim 10^{46-50}$ ergs can be emitted in the form of electromagnetic waves which can be potentially driven by the gravitational field perturbations.

\begin{center}
\begin{figure}[htbp]
\includegraphics[scale=0.5]{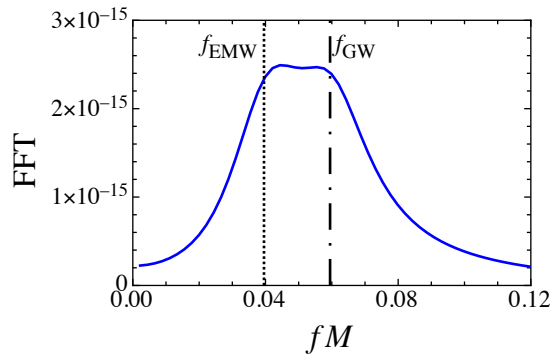}
\caption{
FFT of the electromagnetic waves shown in Fig. \ref{fig:EMW-BH0}. The two vertical lines correspond 
to the frequencies of quasinormal modes for $l=1$ electromagnetic waves (dashed line) and for $l=2$ 
gravitational waves (dot-dash line).
}
\label{fig:PS-BH0}
\end{figure}
\end{center}

Furthermore, in Fig. \ref{fig:PS-BH0}, we show the Fast Fourier Transform (FFT) of the electromagnetic waveforms shown 
in Fig.~\ref{fig:EMW-BH0}, where for comparison we also add the frequencies of the quasinormal modes for $l=1$ electromagnetic 
waves (dashed line) and for $l=2$ gravitational waves (dot-dash line) radiated from the Schwarzschild black hole~\cite{CPM-I}. 
From this figure, one can obviously see that the driven electromagnetic waves have two specific frequencies corresponding to 
the $l=1$ quasinormal mode of electromagnetic waves and the $l=2$ quasinormal mode of gravitational waves. This means that it
might be possible to see the effect of gravitational waves via observation of electromagnetic waves. However, electromagnetic waves 
with such a low frequencies could be coupled/absorbed by the interstellar medium (and/or accretion disk around the central object) 
during the propagation and it will be almost impossible to directly detect the driven electromagnetic waves. The only possible 
way to see the driven electromagnetic waves is the observation of indirect effects, such as synchrotron radiation.

%
%
%
%
%
%
%
%

\subsection{Perturbations on a Neutron-Star Background}
\label{sec:Application2}

In order to examine the coupling between the emitted gravitational and electromagnetic waves in a neutron star background, 
we adopt the same initial conditions as for the black hole case, i.e. the electromagnetic perturbations are set to zero 
and the initial gravitational perturbations are approximated by an ingoing Gaussian wave packet. In the numerical calculations, 
we adopt a grid spacing of $\Delta r=R/200$ and a time step $\Delta t/\Delta r=0.05$, where $R$ is the stellar radius. 
For the background stellar models, we adopt the polytropic equation of state (EOS) of the form $P=K\rho^\Gamma$. 
Then, one can get the waveforms of the reflected gravitational waves and the induced electromagnetic ones.

\begin{figure}[htbp]
\begin{center}
\begin{tabular}{cc}
\includegraphics[scale=0.5]{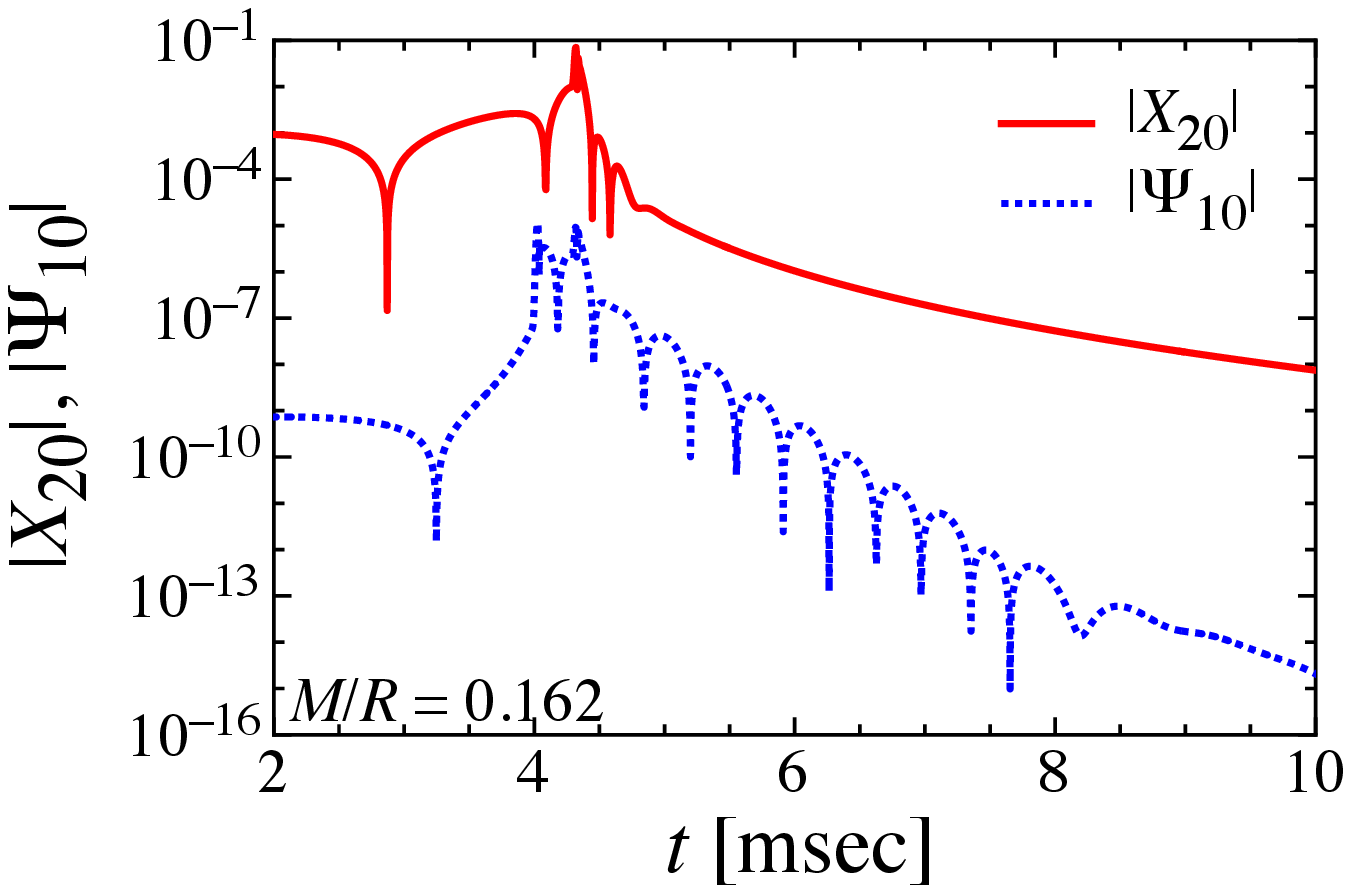} &
\includegraphics[scale=0.5]{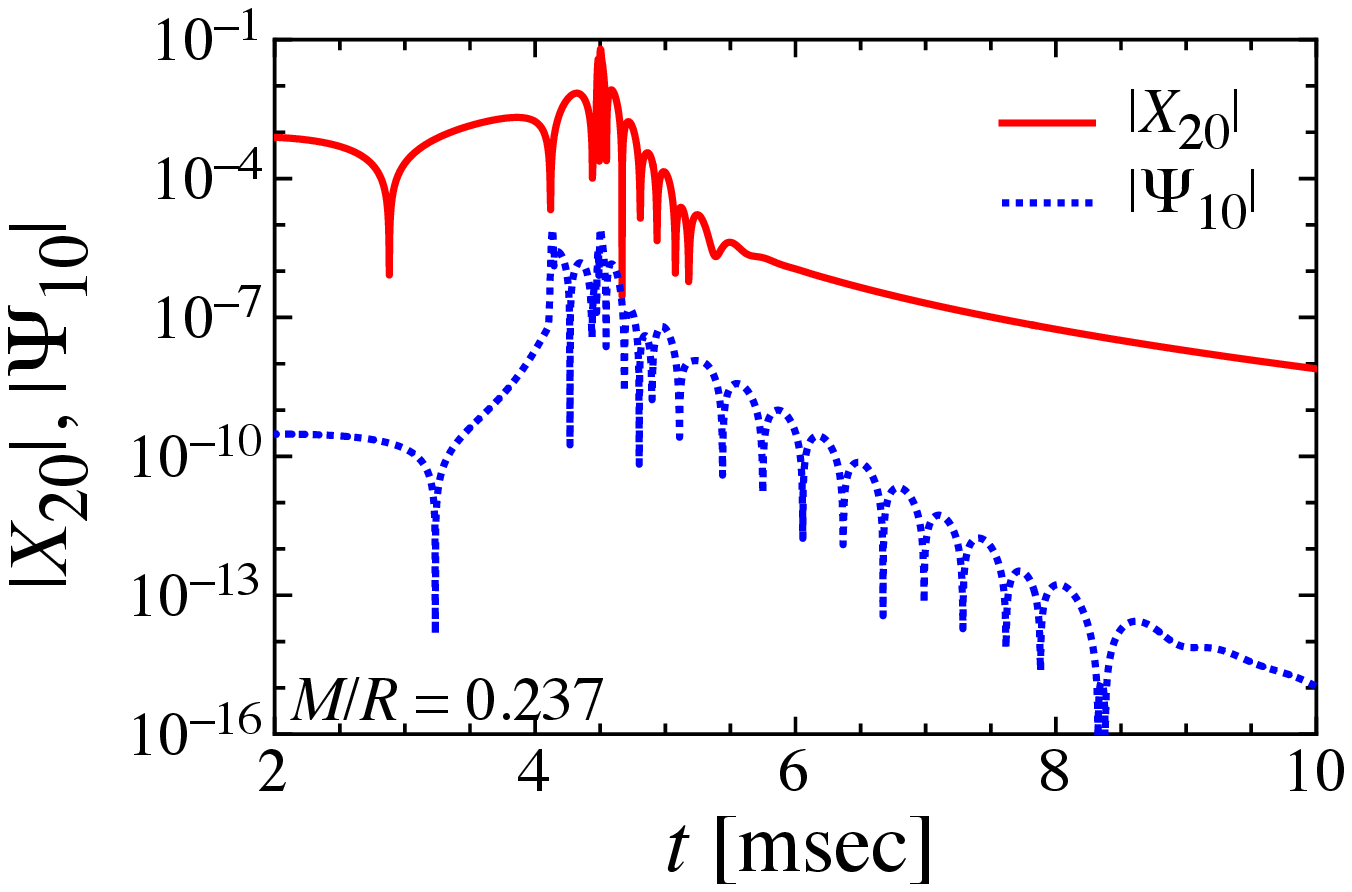} \\
\end{tabular}
\end{center}
\caption{
Waveforms of the gravitational waves (solid line) and the electromagnetic waves (dotted line) for 
the stellar model with $B_p =10^{15}$ Gauss, which are observed at $r=300$ km. The left and right 
panels are corresponding to different stellar models for EOS with $\Gamma=2$ and $K=200$ km$^2$.
}
\label{fig:wave}
\end{figure}

\begin{figure}[htbp]
\begin{center}
\begin{tabular}{cc}
\includegraphics[scale=0.5]{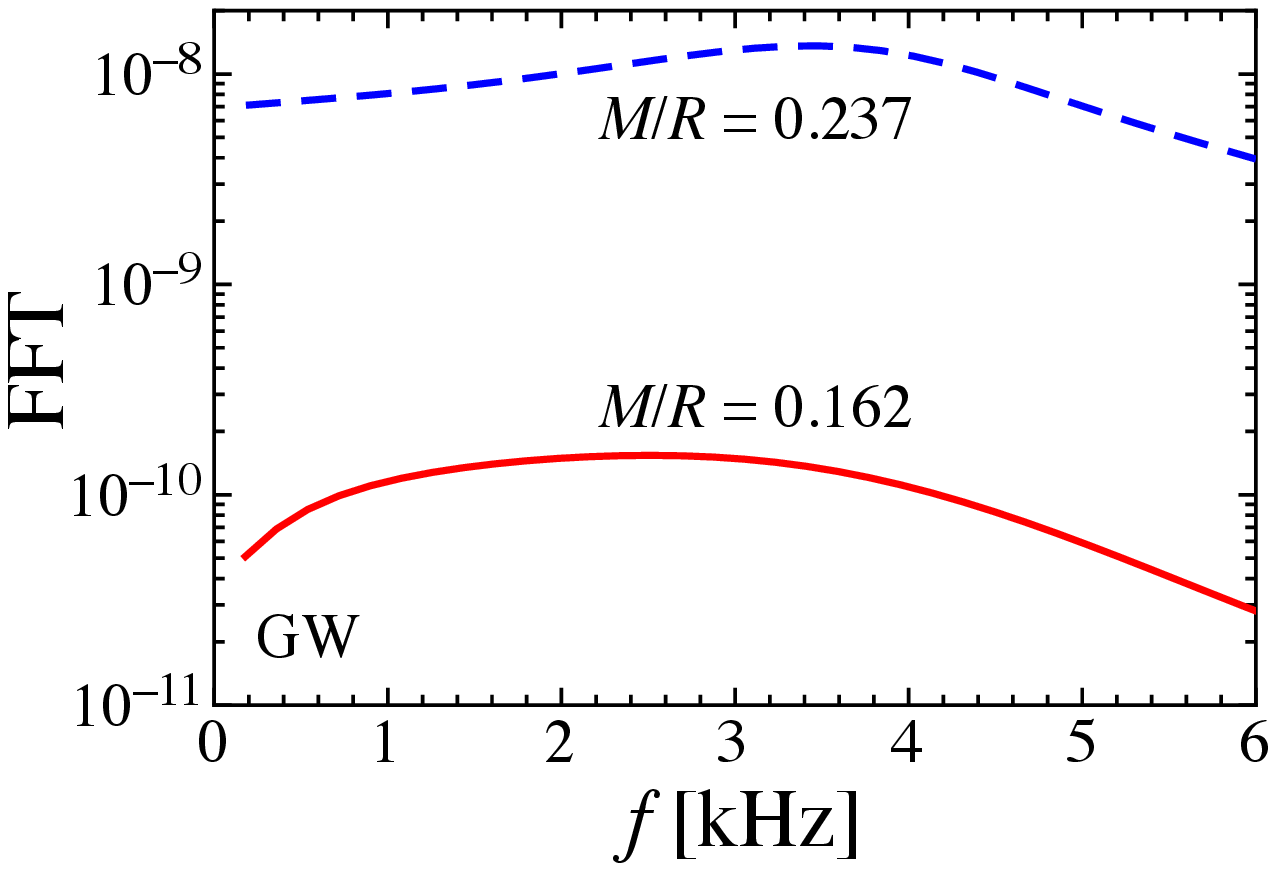} &
\includegraphics[scale=0.5]{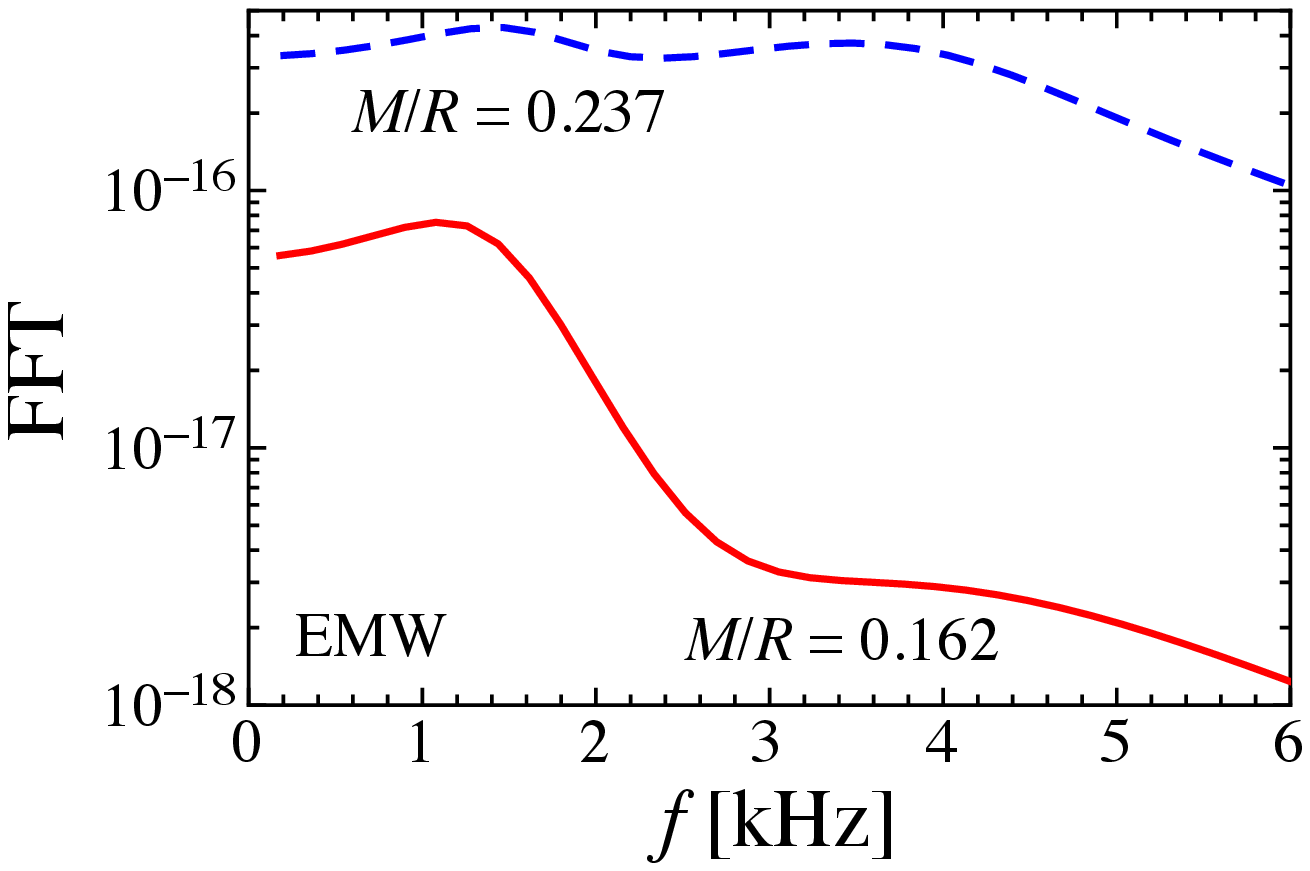} \\
\end{tabular}
\end{center}
\caption{
FFT of the gravitational waves (left panel) and electromagnetic waves (right panel) shown in Fig.~\ref{fig:wave}.
}
\label{fig:PS-NS}
\end{figure}

As an example, we show results for a stellar mode with $\Gamma=2$ and $K=200$ km$^2$. Fig. \ref{fig:wave} shows the 
waveforms of the gravitational waves (solid line) and the electromagnetic waves (dotted line) observed at $r=300$ km, 
where we adopt two stellar models with  different compactness $M/R$  (see Table \ref{Tab:NS} for the stellar properties). 
Compared with the fast damping of gravitational waves, one can see the long-term oscillations in the electromagnetic waves, 
which can be driven not only by the quasinormal ringing of gravitational waves but also during the tail phase of the 
gravitational waves.  For the waveforms shown in Fig.~\ref{fig:wave}, the FFT is plotted in Fig. \ref{fig:PS-NS}, where 
the left and right panels correspond to the FFT of the gravitational and electromagnetic waves, respectively. 
From this figure, one can see the same features as in the case of a black hole. Namely, the FFT of the electromagnetic 
waves driven by the gravitational waves has two specific frequencies, i.e. one is the proper electromagnetic oscillation (1st peak in the right panel of Fig. \ref{fig:PS-NS})
and the other one is the oscillation corresponding to the gravitational waves (2nd peak in the right panel of Fig. \ref{fig:PS-NS}).
We remark that electromagnetic waves with such low frequencies could be absorbed by the interstellar medium and then, their direct detection is almost impossible. Namely, we should consider the secondary emission mechanism such as a synchrotron radiation. Maybe, the plasma around the central object will be excited after receiving the energy from the electromagnetic waves driven by the gravitational waves and move along with the magnetic field lines. Anyway, such a secondary emission mechanism will be discussed somewhere.
Furthermore, we find that as in the case for 
a black hole, the relationship between the emitted energies of gravitational and electromagnetic waves can be described by 
Eq.~(\ref{eq:relation}), even for neutron stars, if $B_p$ is considered as the magnetic field strength at the stellar pole 
($r=R$ and $\theta=0$). In practice, for the specific stellar models in Fig. \ref{fig:wave}, the proportionality constant 
becomes $\alpha = 1.61\times 10^{-5}$ and $4.37\times 10^{-6}$ for the particular stellar models with $M/R=0.162$ and 0.237, 
respectively.

%
%

\begin{center}
\begin{figure}[htbp]
\includegraphics[scale=0.5]{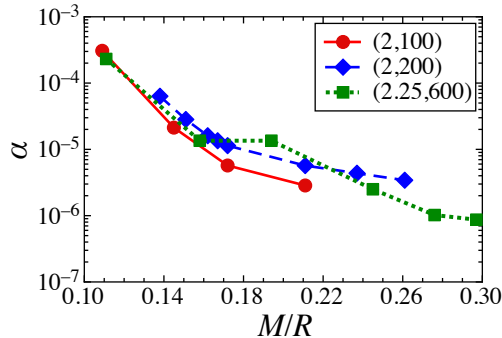}
\caption{
The proportionality constant  $\alpha$ as a function of the stellar compactness for various 
polytropic models. The circles, diamonds, and squares correspond to stellar models with 
$(\Gamma,K)=(2,100), (2,200),$ and $(2.25,600)$.
}
\label{fig:alpha-NS}
\end{figure}
\end{center}

In order to see the dependence on the stellar properties, we study a variety of stellar models with 
different stiffness of the equation of state and with different central densities, radii, and masses, which are 
given in Table~\ref{Tab:NS}. As a result, we find that the proportionality constant $\alpha$ can be written as 
a function of the stellar compactness, which is almost independent of the stellar models and the adopted 
equation of state. In fact, in Fig. \ref{fig:alpha-NS} we plot the values of $\alpha$ for various stellar models, 
where the circles, diamonds, and squares correspond to the results for the stellar models characterized by 
$(\Gamma,K)=(2,100)$, (2,200), and (2.25,600). From this figure, one can see that the proportionality constant 
$\alpha$ depends strongly on the stellar compactness, as expected, with typical values ranging from $10^{-6}$ 
up to $\sim10^{-4}$.

\begin{table}
 \centering
  \caption{Stellar parameters adopted in this article.}
\label{Tab:NS}
  \begin{tabular}{cccccc}
  \hline
   $\Gamma$ & $K$ & $\rho_c$ (g/cm$^3$) & $M/M_\odot$ & $R$ (km) &
   $M/R$ \\
 \hline
 2 & 100 & $1.0\times 10^{15}$ & 0.802 & 10.8 & 0.109  \\
 2 & 100 & $1.5\times 10^{15}$ & 0.998 & 10.2 & 0.145  \\
 2 & 100 & $2.0\times 10^{15}$ & 1.126 & 9.67 & 0.172  \\
 2 & 100 & $3.0\times 10^{15}$ & 1.266 & 8.86 & 0.211  \\
 2 & 200 & $0.7\times 10^{15}$ & 1.365 & 14.6 & 0.138  \\
 2 & 200 & $0.9\times 10^{15}$ & 1.528 & 14.0 & 0.162  \\
 2 & 200 & $1.0\times 10^{15}$ & 1.592 & 13.7 & 0.172  \\
 2 & 200 & $1.5\times 10^{15}$ & 1.791 & 12.5 & 0.211  \\
 2 & 200 & $2.0\times 10^{15}$ & 1.876 & 11.7 & 0.237  \\
 2.25 & 600 & $1.0\times 10^{15}$ & 0.732 & 9.69 & 0.111  \\
 2.25 & 600 & $1.5\times 10^{15}$ & 1.008 & 9.44 & 0.158  \\
 2.25 & 600 & $2.0\times 10^{15}$ & 1.197 & 9.12 & 0.194  \\
 2.25 & 600 & $3.0\times 10^{15}$ & 1.404 & 8.48 & 0.245  \\
 2.25 & 600 & $4.0\times 10^{15}$ & 1.486 & 7.95 & 0.276  \\
\hline
\end{tabular}
\end{table}

%
%

\section{Conclusion}
\label{sec:conclusion}

We have considered the coupling between gravitational and electromagnetic waves emitted 
by compact objects, i.e. black holes and neutron stars. We have derived a coupled system of equations 
describing the propagation of gravitational and electromagnetic waves. In our study we have investigated the 
driving of electromagnetic perturbations via their coupling to the gravitational ones. However, for simplicity, 
we have neglected the back reaction from the electromagnetic waves on the gravitational waves, because the 
magnetic energy of the compact objects, even for magnetars, is quite small as compared with the gravitational 
energy.  We found that the electromagnetic waves of specific parity with harmonic indices $(l,m)$ can be 
coupled to gravitational waves of the same parity and with harmonic indices $(l,m)$ (for $m\ne0$) and 
harmonic indices $(l\pm 1,m)$, for every value of $m$.  In particular, our findings lead to the result that, 
for the axisymmetric perturbations, i.e., $m=0$, the dipole electric electromagnetic waves will be driven by 
axial quadrupole gravitational waves, while the dipole magnetic electromagnetic waves will be driven by polar gravitational waves.

As an application of our perturbative framework, we presented numerical calculations for the case in which 
dipole-electric electromagnetic waves are driven by the axial gravitational ones, both for the case of a 
black hole and a neutron star background. We found that the emitted energy in electromagnetic 
waves driven by the gravitational waves is proportional to not only the emitted energy in gravitational 
waves but also to the square of the strength of the magnetic field of the central object. For the case of a 
black hole background, the ratio of the emitted energy of the electromagnetic waves to that of the 
gravitational waves is around $8\times 10^{-6}(B_p/10^{15} {\rm G})^2$, where $B_p$ is the magnetic field 
strength at $r=6M$. On the other hand, in the case of a neutron star background, we find that this 
proportionality constant can be written as a function of the stellar compactness. 

Although we have considered only the case of axial gravitational waves and the associated induced 
electromagnetic waves, the polar oscillations also play an important role in extracting the information 
about the neutron star structure since in the case of non-rotating stars, the matter oscillations are 
typically coupled to the polar gravitational waves. This is a direction that we are currently investigating.

\acknowledgments

H.S. is grateful to Ken Ohsuga for valuable comments. This work was supported by the German Science Foundation (DFG) via SFB/TR7, by Grants-in-Aid for Scientific Research on Innovative Areas through No.\ 23105711, No.\ 24105001, and No.\ 24105008 
provided by MEXT, by Grant-in-Aid for Young Scientists (B) through No.\ 24740177 provided by JSPS,
by the Yukawa International Program for Quark-hadron Sciences,
and by the Grant-in-Aid for the global COE program ``The Next Generation of Physics, Spun from Universality and Emergence" from MEXT.
C.F.S. acknowledges support from contracts FIS2008-06078-C03-03, AYA-2010-15709, and FIS2011-30145-C03-03 of 
the Spanish Ministry of Science and Innovation, and contract 2009-SGR-935 of AGAUR (Generalitat de Catalunya).
P.L. acknowledges the support from NSF awards 1205864, 0903973 and 0941417.



\end{document}